\documentclass{emulateapj}

\usepackage{psfig}
\usepackage{amsmath}
\usepackage{amssymb}
\usepackage{graphicx}

\newcommand{\FLASH}{{\sc flash}}

\shortauthors{Zhang {\it et al.}}
\shorttitle{Evolution of Thermonuclear Flames} 

\begin{document}

\title{On the Evolution of Thermonuclear Flames on Large Scales}

\author{Ju Zhang\altaffilmark{1,2},
  O. E. Bronson Messer\altaffilmark{1,2,3}, Alexei M.
  Khokhlov\altaffilmark{1,2} and Tomasz Plewa\altaffilmark{1,2}}
\altaffiltext{1}{Center for Astrophysical Thermonuclear Flashes,
   The University of Chicago,
   Chicago, IL 60637}
\altaffiltext{2}{Department of Astronomy \& Astrophysics,
   The University of Chicago,
   Chicago, IL 60637}
\altaffiltext{3}{present address:  National Center for Computational Sciences,
        Oak Ridge National Laboratory, Oak Ridge, TN 37831}
        
\begin{abstract} 

The thermonuclear explosion of a massive white dwarf in a Type Ia
supernova explosion is characterized by vastly disparate spatial and
temporal scales. The extreme dynamic range inherent to the problem
prevents the use of direct numerical simulation and forces modelers
to resort to subgrid models to describe physical processes
taking place on unresolved scales.

We consider the evolution of a model thermonuclear flame in a constant
gravitational field on a periodic domain. The gravitational
acceleration is aligned with the overall direction of the flame
propagation, making the flame surface subject to the
Rayleigh-Taylor instability. The flame evolution is followed through
an extended initial transient phase well into the steady-state
regime. The properties of the evolution of flame surface are examined.
We confirm the form of the governing equation of the
evolution suggested by Khokhlov (1995). 

The mechanism of vorticity production and the interaction between
vortices and the flame surface are discussed. Previously observed
periodic behavior of the flame evolution is reproduced and is found to
be caused by the turnover of the largest eddies. The characteristic
timescales are found to be similar to the turnover time of these
eddies. Relations between flame surface creation and destruction
processes and basic characteristics of the flow are discussed. We
find that the flame surface creation strength is associated with the
Rayleigh-Taylor time scale. Also, in fully-developed
turbulence, the flame surface destruction strength scales as $1/L^3$,
where $L$ is the turbulent driving scale.

The results of our investigation provide support for Khokhlov's
self-regulating model of turbulent thermonuclear flames (Khokhlov
1995).  Based on these results, one can revise and extend the original
model. The revision uses a local description of the flame surface
enhancement and the evolution of the flame surface since the onset of
turbulence, rendering it free from the assumption of an 
instantaneous steady state of turbulence. This new model can be applied
to the initial transient phase of the flame evolution, where the self-regulation mechanism yet to be 
fully established. Details of this new model will be presented in a forthcoming paper. 

\end{abstract}

\section{Introduction}\label{s:intro}
Thermonuclear combustion was originally proposed as the mechanism
behind Type Ia supernovae \citep{hoyle+60,arnett69}.
However, this detonation-based model was soon shown to be unable to explain a host
of observations \citep{arnett74,ostriker+74}. \citet{nomoto+76}
recognized the possibility that a deflagration, a subsonic mode of
combustion following a mild ignition of stellar material in the center
of a massive white dwarf, might help to ameliorate these deficiencies.
Flames characteristic of a deflagration are subject to the
Rayleigh-Taylor instability in the gravitational field in the interior
of a white dwarf, significantly complicating any detailed calculation of 
this mode of burning.

The parameterized one-dimensional
study of \cite{nomoto+84} indicated that a successful explosion model
may require a substantial increase in the speed of the deflagrating front.
Such an increase in burning speed was observed in the
multi-dimensional models of \cite{mueller+82,mueller+86}, though  \cite{mueller+86} recognized that their
models lacked a proper description of turbulence (and also lacked details of the 
microphysics) taking place on scales unresolved in their
simulations.

The structure and dynamics of deflagration fronts on
Rayleigh-Taylor-dominated scales were studied in detail by
\cite{khokhlov95}. In particular, Khokhlov proposed that when a 
steady state of fully-devloped turbulence is assumed, the evolution of the flame propagation becomes
independent of details of the microphysics and is governed by a
self-regulating mechanism reflecting an interplay between the
Rayleigh-Taylor instability and nuclear burning. Khokhlov (1995) provides 
scaling relations for the turbulent flame speed and these scalings are used  
to construct a subgrid-scale (SGS) model which, combined with
a flame capturing scheme, offers a powerful tool for studying
deflagrations in Type Ia supernovae.

The use of subgrid-scale models in modeling of Type Ia supernovae is
necessitated by the vastly disparate temporal and spatial scales
present in the problem. Without exception, all groups involved in
studying thermonuclear supernovae use some sort of subgrid-scale
model. In the Khokhlov self-regulation scenario, the turbulent flame
speed is primarily determined by the speed of large-scale motions
driven by the Rayleigh-Taylor instability. This speed is $\propto
\sqrt{AgL}$, where $A$ is the Atwood number, $g$ is the local
gravitational acceleration, and $L$ is the driving scale.  In this
subgrid model, it is assumed that a steady state of fully-developed
turbulence obtains instantaneously, leading to a state of self
regulation where the interplay of the RT instability and the local
progress of the flame serves to render the turbulent flame speed
roughly constant in time.  However, the model does not take into
account the initial transient phase nor the unsteadiness of local
turbulent velocities.  The subgrid model of \cite{niemeyer+95} uses
the framework of Kolmogorov turbulence theory as a starting point. In
their model, the burning speed is locally determined from the
turbulent kinetic energy, $k_{sgs}$, which is evolved through the use
of a local transport equation \citep{clement93}. This model has been
used extensively to study thermonuclear deflagrations both without
\citep{niemeyer+96} and, later, with \citep{reinecke+99,reinecke+02} a
flame front tracking scheme. Despite the fundamental differences in
these subgrid models, the results obtained
(explosion energetics, timescale, and ejecta morphology) are
quite similar for deflagrations in the self-regulation regime
\citep{reinecke+99,reinecke+02,gamezo+03,Schmidt+06}.  Accordingly, we can
conclude that in the self-regulation regime differences between
specific subgrid models do not lead to gross differences in the flame
evolution.

Our aim here is to improve the original Khokhlov subgrid-scale model
(hereafter KSGS) and develop a new subgrid model that can be applied
not only to the self-regulation regime but also to the initial
transient phase before self regulation has been fully established. The
inclusion of a description of this transient phase in the subgrid
model is especially important for the simulation of the early
evolution of deflagrations in Type Ia supernovae.

To develop the new subgrid model, we have first conducted a survey of
thermonuclear model flame simulations on periodic domains. These
results allow us to develop a deeper understanding of the flame
dynamics on small scales and provide a foundation for the improved
scheme. 

\section{Numerical methods and models}
In our study of model thermonuclear flames, we consider the evolution
of the reactive Eulerian equations  in Cartesian
geometry. Numerical solutions were obtained with the \FLASH\ code
\citep{fryxell+00}. The majority of the models presented here were computed in
three-dimensions, while a limited number of models have been
calculated in two-dimensions.
\subsection{Numerical Methods}\label{s:methods}
We assume a constant gravitational acceleration, $g_z$, aligned with
and pointing along the negative z coordinate. The computational domain
is a tube (a channel in 2D) with square cross-section. We use periodic
boundary conditions along the lateral (i.e.\ $(x,y)$) directions. We impose
reflecting boundary conditions at the bottom of the tube, while
allowing outflow through its top boundary.  The time steps in our
simulation were limited with Courant factor of 0.6.

In order to save computational resources, we use {\FLASH}'s adaptive
mesh refinement (AMR) capability and only resolve regions actively
participating in combustion. In these regions, we track flow
structures characterized by large jumps in density, total
velocity, or progress variable.

At the initial time, a one-dimensional flame front structure is
interpolated onto the grid with the flame front position, $z_f$, defined
as

\[
   z_f = z_0 + A_f \cos( 2 \pi x / \lambda_x ) \cos( 2 \pi y / \lambda_y )\,,
\]

where $z_0$ is the unperturbed position of the front, $A_f$ is the
amplitude of the perturbation, and $\lambda_x =\lambda_y = \lambda$ is
the wavelength of the perturbation in the lateral directions. The fuel filling
the tube ahead of the flame front is initially at rest and composed entirely
of equal mass fractions of carbon and oxygen. The fuel's
density is $1\times 10^8$ g/cm$^3$ and its temperature is
$5\times 10^7 K$. We consider here only a one-step reaction in the burning front with a
constant prescribed energy release of $7\times 10^{17}$ erg/g (i.e. the total energy that
would be released in taking the initial carbon-oxygen mix to a composition of all iron). 
We use a Helmholtz equation of state
\citep{timmes+00} suitable
for the degenerate conditions considered here.

The original \FLASH\ code has been equipped with a custom
implementation of the flame-capturing algorithm of
\citet{khokhlov95,gamezo+03}. In this framework, the
flame front is described as a diffuse, thick interface of a progress
variable, $f$, whose evolution is governed by an
advection-diffusion-reaction (ADR) equation:

\begin{equation} \label{e:adr}
  \frac{\partial f}{\partial t} + U\cdot\nabla{f} = K\nabla^2{f} + R\,,
\end{equation}

Here $f$ is constrained to lie between $f = 0$ (unburned material, or fuel) and
$f = 1$ (completely burned material, or ash).  The diffusion and
reaction coefficients, $K$ and $R$, are given by

\[
\begin{array}{ll}
  K = \mathrm{const}, & \\
  R = \left\{
      \begin{array}{ll}
              C = \mathrm{const}, & \mathrm{if}\ f_0 \leq f \leq 1\,, \\
              0,                  & \mathrm{otherwise}\,, \\
      \end{array}
      \right.
\end{array}
\]
 
where $f_0 = 0.3$. Equation~(\ref{e:adr}) describes a reactive front
of thickness $\delta \simeq (K/C)^{1/2}$ propagating with the speed $D
= (KC/f_0)^{1/2}$. With coefficients $K$ and $C$ defined as

\[
  K = D(\beta\Delta x)\sqrt{f_0},\ C = \frac{D}{(\beta\Delta x)}\sqrt{f_0},
\]

and $\beta = 1.5$, the front diffuses with a prescribed constant speed
$D$ and spreads over $\approx 3-4$ zones. For the advection part of
the ADR scheme we use the PPM \citep{colella+84} solver in
{\FLASH}. For a more detailed description of the flame-capturing
method and its characteristics see \citet{khokhlov95} and
\citet{zhiglo05}. Verification of the \FLASH\ implementation of this
scheme has been presented in \cite{vladimirova+05}.

\subsection{Database of Numerical Models}
In order to gain insight into the nature of the evolution of
turbulent thermonuclear flames on large scales, we have constructed an
extensive database of numerical flame models.  

\begin{table*}
\caption{Model Turbulent Thermonuclear Flames}\label{t:dns_runs} 
\begin{tabular}{ll}
Parameter&Definition\\
\hline
$N_x \times N_y \times N_z$ & model resolution \tablenotemark{a}\\
$L$                         & lateral extent of the computational domain\\
$D_l$                       & laminar flame speed\\
$D_t$                       & turbulent flame speed\\  
$g_z$                       & gravitational acceleration\\
$Fr$                        & Froude number, $Fr = D_l^2/g_zL$\\
$A_f$                       & amplitude of the flame front perturbation in vertical direction \\
$\lambda$                   & wavelength of the flame front perturbation in lateral directions \\
$b$                         & nominal numerical thickness of the flame
front (number of computational cells) \\
\end{tabular}
\tablenotetext{a}{Effective resolution of AMR model.}
\end{table*}

\begin{table*}
\caption{High-Resolution non-SGS Models\tablenotemark{a}}
\begin{tabular}{cccccccc}
Model & $N_x \times N_y \times N_z$ & $L$ [cm] & $\Delta x$ [cm] & $D_l$ [cm s$^{-1}$] & $g_z$ [cm s$^{-2}$] & $Fr\times 10^4$ & $b$ \\
\hline
R1 & $ 64  \times  64 \times  2560$ & $1.5  \times 10^6$ & $2.34\times 10^4$ & $1.07 \times 10^6$ & $1.9 \times 10^9$ & 4.02 & 4\\
R2 & $128  \times 128 \times  5120$ & $1.5  \times 10^6$ & $1.17\times 10^4$ & $1.07 \times 10^6$ & $1.9 \times 10^9$ & 4.02 & 4\\
R3 & $256  \times 256 \times 10240$ & $1.5  \times 10^6$ & $5.86\times 10^3$ & $1.07 \times 10^6$ & $1.9 \times 10^9$ & 4.02 & 4\\
BH & $128  \times 128 \times  5120$ & $1.5  \times 10^6$ & $1.17\times 10^4$ & $1.07 \times 10^6$ & $1.9 \times 10^9$ & 4.02 & 3\\
B2 & $128  \times 128 \times  5120$ & $1.5  \times 10^6$ & $1.17\times 10^4$ & $1.07 \times 10^6$ & $1.9 \times 10^9$ & 4.02 & 6\\
DH & $ 64  \times  64 \times  2560$ & $1.5  \times 10^6$ & $2.34\times 10^4$ & $0.54 \times 10^6$ & $1.9 \times 10^9$ & 1.01 & 4\\
D2 & $ 64  \times  64 \times  2560$ & $1.5  \times 10^6$ & $2.34\times 10^4$ & $2.14 \times 10^6$ & $1.9 \times 10^9$ & 16.1 & 4\\
LH & $ 64  \times  64 \times  2560$ & $0.75 \times 10^6$ & $2.34\times 10^4$ & $1.07 \times 10^6$ & $1.9 \times 10^9$ & 8.04 & 4\\
L2 & $128  \times 128 \times  5120$ & $3.0  \times 10^6$ & $1.17\times 10^4$ & $1.07 \times 10^6$ & $1.9 \times 10^9$ & 2.01 & 4\\
GH & $ 64  \times  64 \times  2560$ & $1.5  \times 10^6$ & $2.34\times 10^4$ & $1.07 \times 10^6$ & $9.5 \times 10^8$ & 8.04 & 4\\
\end{tabular}
\tablenotetext{a}{$A_f = 9.96 \times 10^4 cm$, $\lambda = 4.46 \times 10^5 cm$.}
\end{table*}

Tables~\ref{t:dns_runs} and 2 provides a complete list of the catalogued simulations.

Our high resolution models are primarily intended to extend the
simulations of \citet{khokhlov95}, verify our implementation of the
flame capturing algorithm, verify the KSGS model, and demonstrate
numerical convergence. Specifically, models R1, R2 and R3 are used to
investigate grid convergence. Models BH and B2 together with model R2
are used to demonstrate the insensitivity of the results to the
numerical flame thicknesses, $b$. Models DH, R1 and D2 are used in
a discussion of flame dynamics, including the self-regulation
mechanism. In models LH, R1 and L2 we vary the size of computational
domain.  These models are analized in our study of the
flame destruction process. Models R1, D2, GH and L2, having various
domain sizes and the imposed 
gravitational accelerations, are used to verify
the KSGS model. 
\section{Flame Dynamics}\label{s:dynamics}
Following \citet[see also
\citet{khokhlov95,niemeyer+95}]{damkoehler39}, a primary consequence
of turbulence is to increase the flame surface area. The turbulent
flame speed is,

\[
D_t = \rho_0^{-1} \frac{dM_b}{dt}\,,
\]

where the burned mass, $M_b$, is normalized to the area of the unperturbed flame the horizontal cross-section of the computational domain.  
The turbulent flame speed is expected to vary in proportion to the surface area,
$S$, as

\begin{equation} \label{e:Dt_S}
  D_t = \frac{S}{S_0}D_l\,, 
\end{equation}

where $S_0$ is the surface area of an unperturbed flame.
\citet{khokhlov95} suggested, based on examination of  simple flame surface
geometries, that the evolution of the flame surface area can be described
as a competition between flame surface creation and destruction
processes:  

\begin{equation} \label{e:dSdt}
  \frac{dS}{dt} = cS - dD_lS^2\,,
\end{equation}

where

\begin{equation} \label{e:creation}
  c = [(\mathbf{e}_1)_i(\mathbf{e}_1)_j + (\mathbf{e}_2)_i(\mathbf{e}_2)_j](\frac{\partial U_i}{\partial x_j}).
\end{equation}

Here $(\mathbf{e}_{1,2})_{i}$ are two orthonormal vectors tangent
to an infinitesimal surface element $\delta S$. The first unit vector,
$\mathbf{e}_1$, is orthogonal to vector normal to the flame
surface, $\mathbf{n} = \nabla f/|\nabla f|$, and is chosen
arbitrarily; the second unit vector, $\mathbf{e}_2$, is orthogonal
to both $\mathbf{e}_1$ and $\mathbf{n}$ and is then determined
uniquely.

The first term on the right-hand side of Eq.~(\ref{e:dSdt}) describes
surface creation by strain, while the second term in this equation represents the
process of flame surface destruction due to the propagation of cusps. In
steady state, the two terms balance one another in a
statistical sense. Then, $S \propto D_l^{-1}$ follows from
Eq.~(\ref{e:dSdt}) implying that the turbulent flame speed $D_t$ is
independent of the laminar flame speed, $D_l$.

Each term in Eq.~(\ref{e:dSdt}) can be calculated using available
information from our simulations.  The flame surface area, $S$, is
extracted from the data as an iso-surface of the progress variable at
$f = 0.5$ using a marching-cube algorithm \citep{mcube}. The
computational cells that contain flame surface (defined by $f = 0.5$)
are then used to calculate the surface creation coefficient, $c$:
First, a unit vector normal to the flame surface, $\mathbf{n}$, is
calculated. Then two unit vectors, $\mathbf{e}_1$ and $\mathbf{e}_2$,
orthogonal to $\mathbf{n}$ and to each other are constructed
\citep{khokhlov95}. It is the average value of the creation
coefficient that is followed, $\int_\Omega cdV/\int_\Omega dV$, where
$\Omega$ is the volume containing flame surface, {\it i. e.} all of
the computational cells that contain flame surface. The total flame
surface area, $S$, and creation term, $cS$, are computed by
integrating over the entire flame surface with $S = \int_\Omega dS$
and $cS = \int_\Omega cdS$. The total destruction term is then
calculated as the difference between total creation and $dS/dt$, where
$dS/dt$ is simply calculated by finite differencing in time.

\subsection{Flame Surface Evolution}\label{s:S_evol}
\begin{figure}[h]
\begin{center}
\includegraphics[height=7.cm,clip=true]{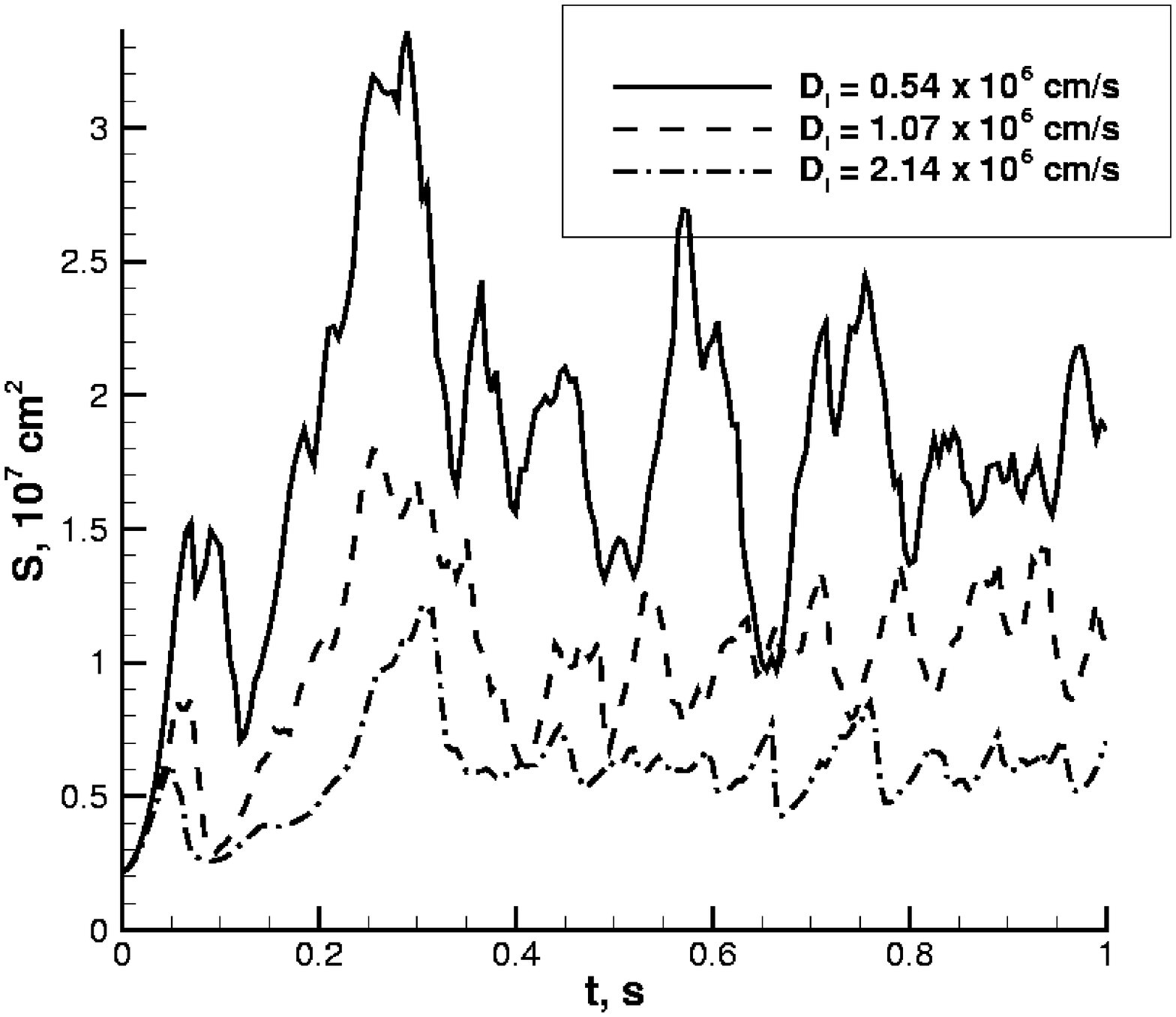}
\includegraphics[height=7.cm,clip=true]{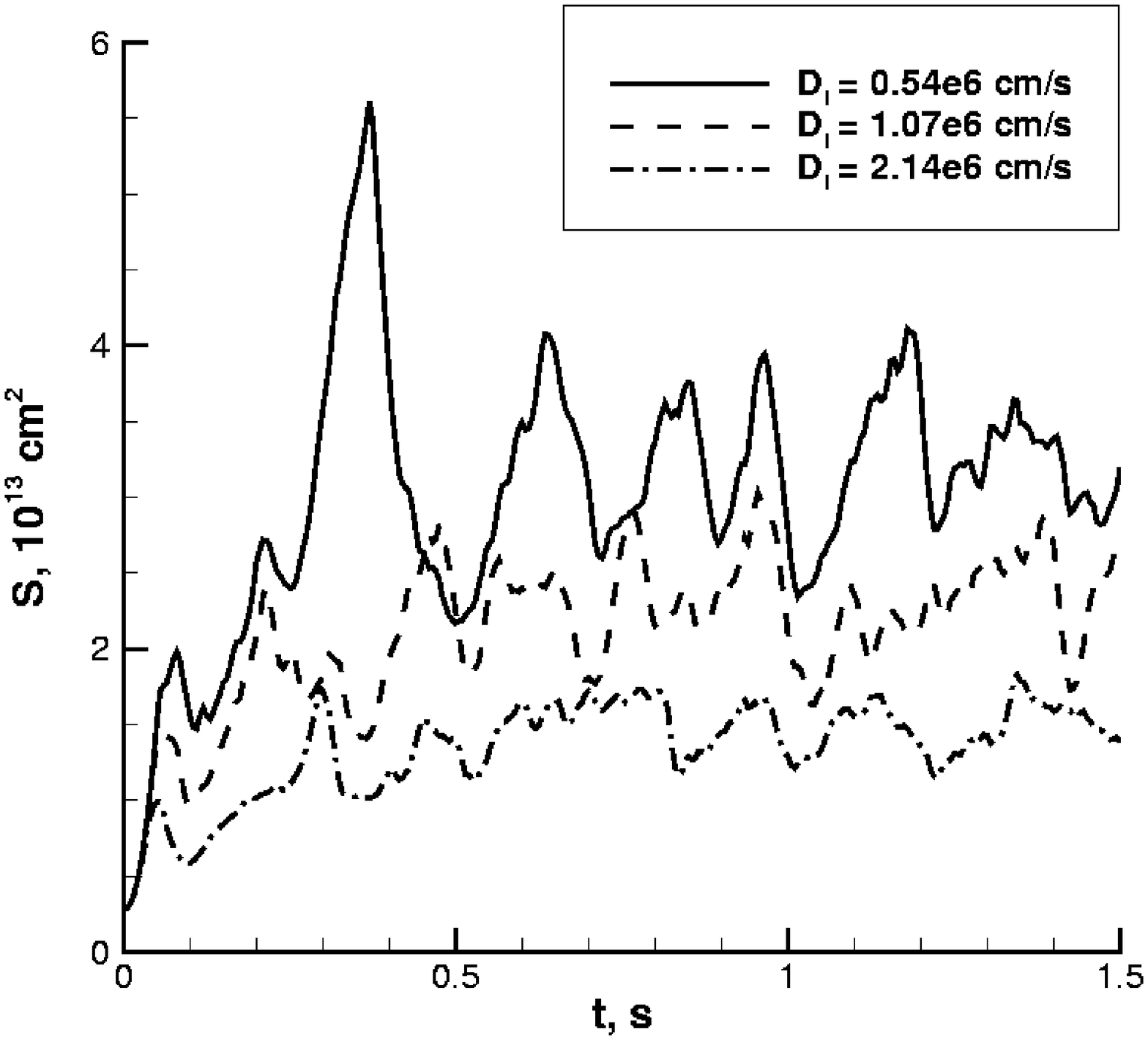}
\end{center}
\caption{Temporal evolution of flame surface area in 2D (top) and 3D
  (bottom). Note that in 2D, the models are the counterparts of models
  DH, R1 and D2 in 3D.} 
\label{f:S_Dl}
\end{figure}

We begin with an examination of the evolution of the model flames listed
in Tables~\ref{t:dns_runs} and 2. These results serve as the
basis for the analysis of the actual form of the governing equation
for flame surface evolution. 
First, we examine the evolution of models with different laminar flame speeds,
$D_l$. These results are closely associated with the
self-regulation mechanism that will be discussed throughout this
paper. Fig.~\ref{f:S_Dl}

shows the evolution of the flame surface area at three different laminar
flame speeds both in 2D and 3D. It is expected that the larger
the laminar flame speed, the smaller the flame surface.  As discussed
by \cite{khokhlov95} and shown in Sec.~\ref{s:Fr}, the
larger the laminar flame speed, the stronger the smoothing effect due
to burning, thus the smaller the flame surface area.

Even a crude inspection of Fig.~\ref{f:S_Dl} suggests that a steady
state exists and the flame exhibits a semi-periodic behavior about
this equilibrium.\footnote{The character of the steady state will be
further elaborated in Sec.~\ref{s:steady}.}  The time averaged value
and the standard deviation of the flame surface area in steady state
(at t $>$ 0.6 s) are listed in Table~\ref{t:S_mean}.

\begin{table}
\caption{Statistics of Flame Surface in Steady State}\label{t:S_mean} 
\begin{tabular}{ccc}
$D_l$ $\times 1.07 \times 10^{-6}$ cm/s (2D) & $\bar{S}$ $\times 10^{-7}$ cm$^2$ & $S'$
  $\times 10^{-7}$ cm$^2$\\
\hline
0.5& 1.8 & 0.9 \\
1.& 1. & 0.5 \\
2.& 0.5 & 0.3\\

\hline
\hline
$D_l$ $\times 1.07 \times 10^{-6}$ cm/s (3D) & $\bar{S}$ $\times 10^{-13}$ cm$^2$ & $S'$
$\times 10^{-13}$ cm$^2$\\
\hline
0.5. & 3.2 & 1.2 \\
1.   & 2.5 & 1   \\
2.   & 1.2 & 0.5 \\

\end{tabular}
\end{table}

\begin{figure}[ht]
\begin{center}
\includegraphics[height=7.cm,clip=true]{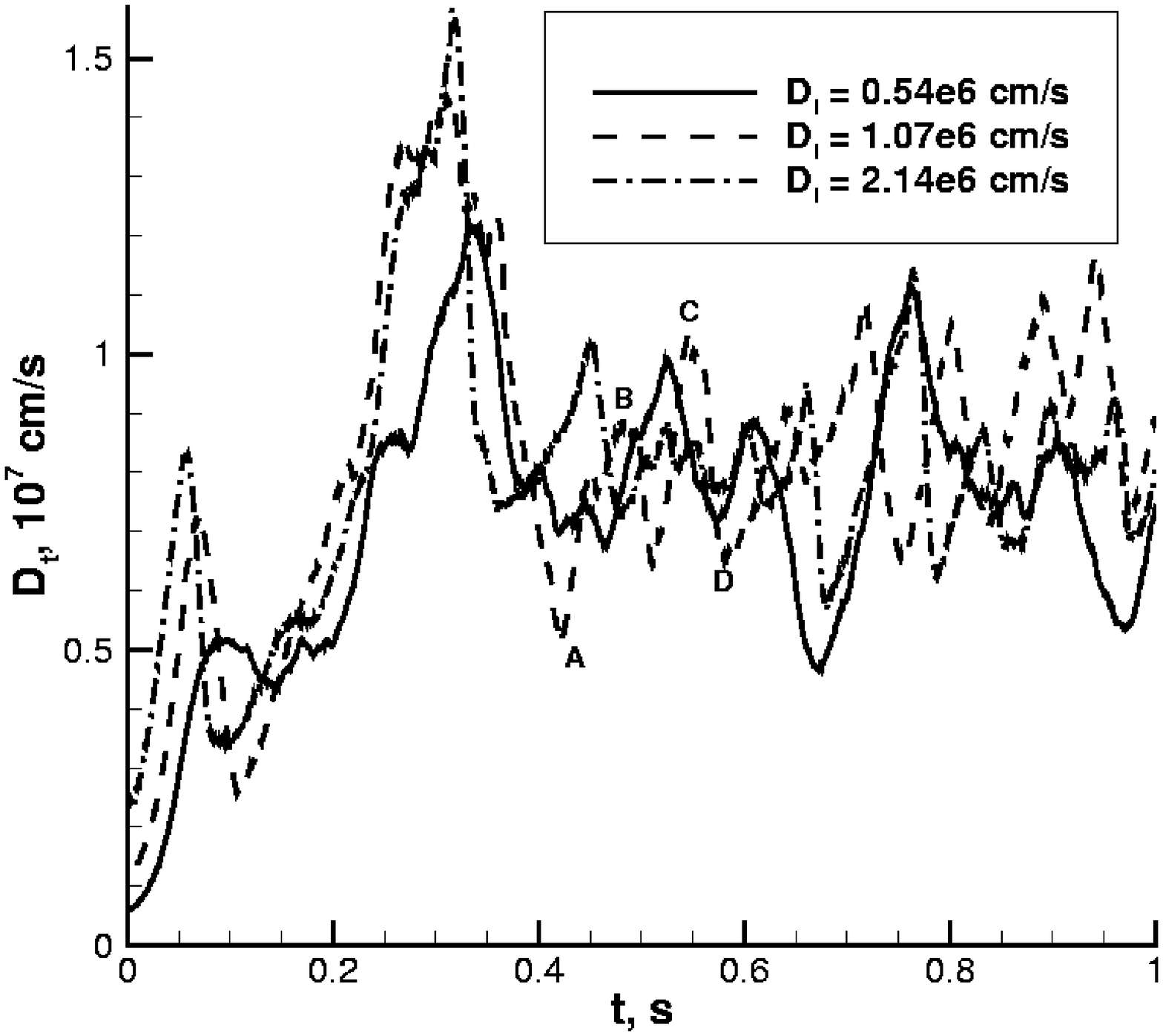}
\includegraphics[height=7.cm,clip=true]{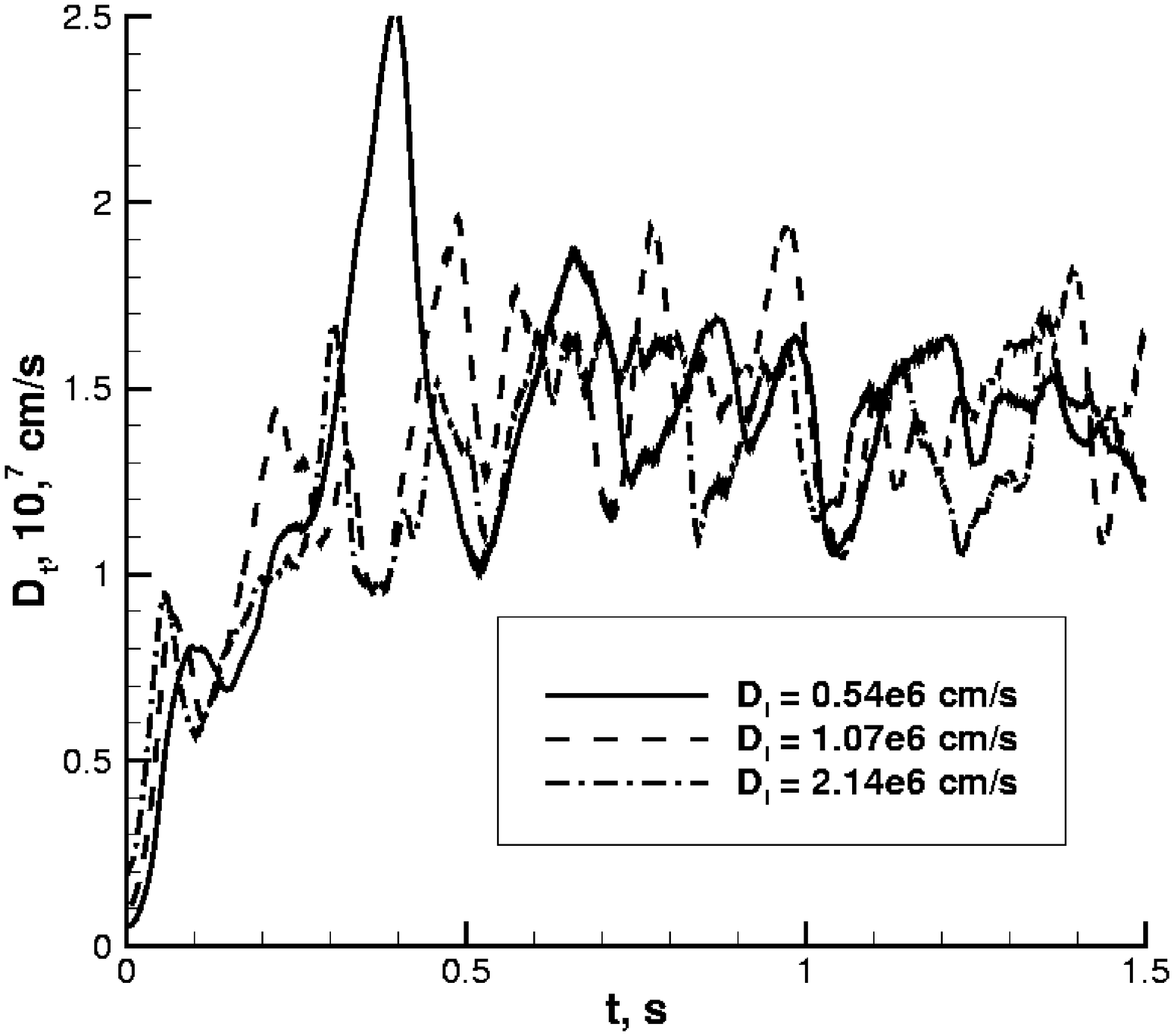}
\end{center}
\caption{Evolution of the turbulent flame speed, $D_t$ in 2D (left) and 3D (right) in models DH (solid
  line), R1 (dashed line) and D2 (dash-dotted line). Note that a
  statistically steady-state evolution is preceded by an extended
  transient. The time-averaged turbulent flame speed does not depend
  on the laminar flame speed once a steady-state regime is established, 
  where the evolution is governed by the self-regulation of the flame surface.}
\label{f:Dtself}
\end{figure}

%
%
%
Both the time averaged value and the standard deviation of the
flame surface area appear to scale as $D_l^{-1}$. In most cases
considered this relation is fulfilled with accuracy of $\approx
10 \%$. Following our trial time-limited high-resolution runs, this
agreement seems improving with increased resolution. Both the time
averaged turbulent flame speed and its deviation are independent of
the laminar flame speed. This is in agreement with Eq.~(\ref{e:Dt_S})
and is illustrated in Fig.~\ref{f:Dtself}.

Therefore, our study confirms the independence of the
turbulent flame speed and the laminar flame speed.  This is  the essential property of the
self-regulation mechanism discussed by \cite{khokhlov95}.
However, the independence of the deviation of turbulent flame
speed versus laminar flame speed has not been discussed in that
previous study.
%
%
%
\subsection{Character of The Steady State}\label{s:steady}
As discussed in the previous section,  the flame exhibits a
semi-periodic behavior once a steady state is established. A natural question then
is ``What is the source of the oscillation?'' It is conceivable that
the oscillation is caused by interference between the flame and
acoustic waves generated at the bottom wall boundary.  On the other hand, it could be an intrinsic property of Rayleigh-Taylor unstable
flames. In order to determine whether of these hypotheses is likely we perform a 1D test without flame propagation. In
this test an initial vertical downward velocity is imposed on the
fluid and the propagation of sound waves generated at the bottom
wall is examined. We find that the sound waves quickly reach
the top boundary and leave the computational domain. The fluid
becomes essentially at rest once the sound wave propagates off the grid. In addition,
a simulation of flame propagation in 2D with both the top and bottom
boundaries open have been performed and the same semi-periodic behavior found in our
fiducial models  is observed. 

\begin{figure}[ht]
\begin{center}
\includegraphics[height=5.4cm,clip=true]{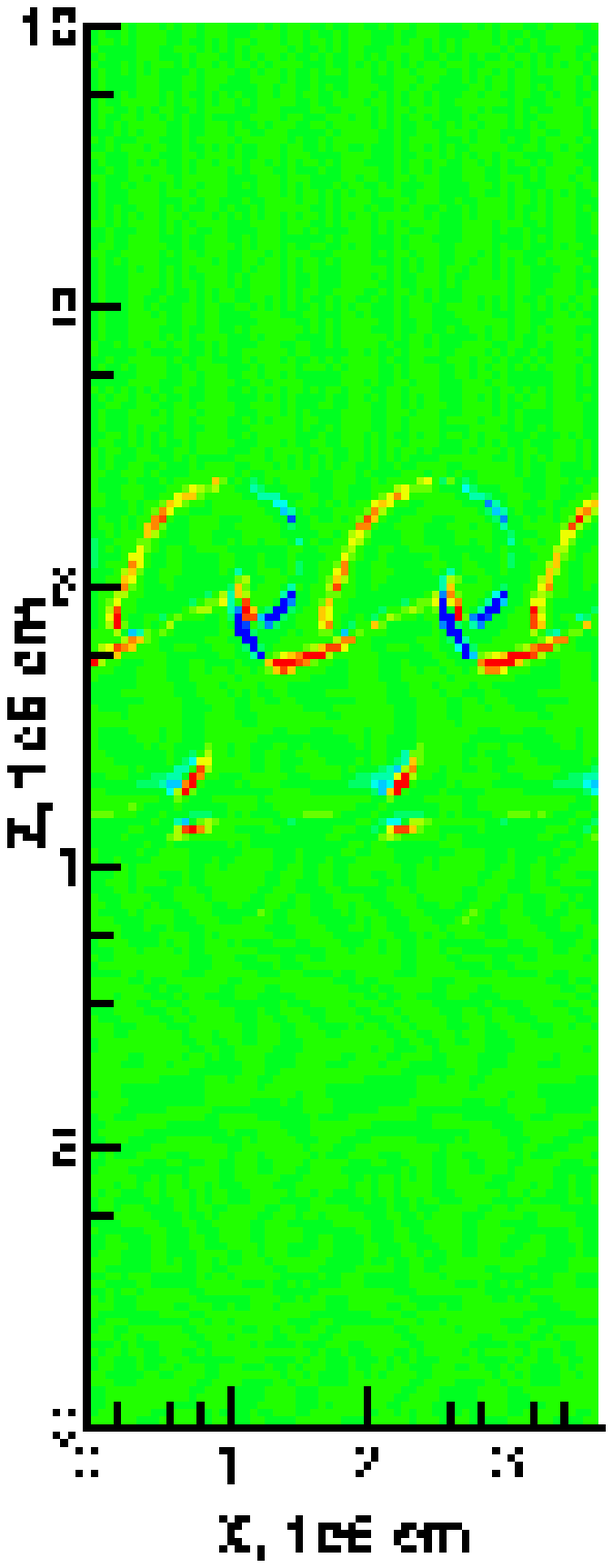}
\includegraphics[height=5.4cm,clip=true]{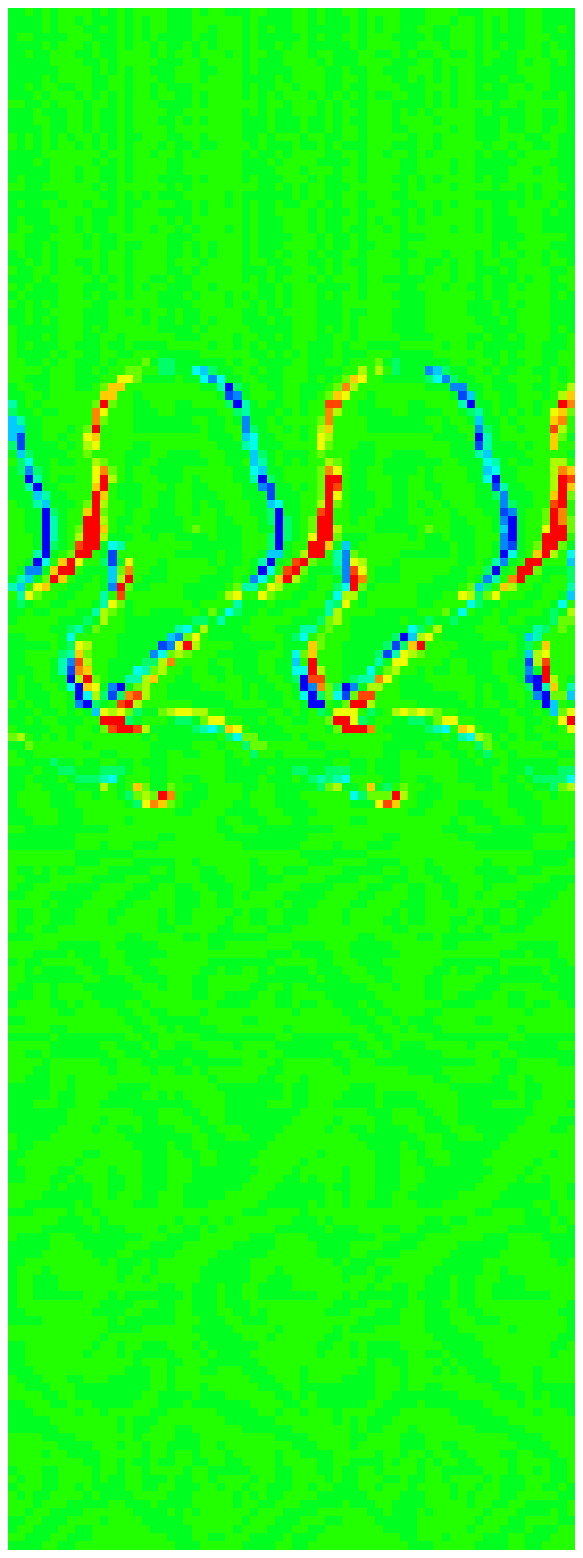}
\includegraphics[height=5.4cm,clip=true]{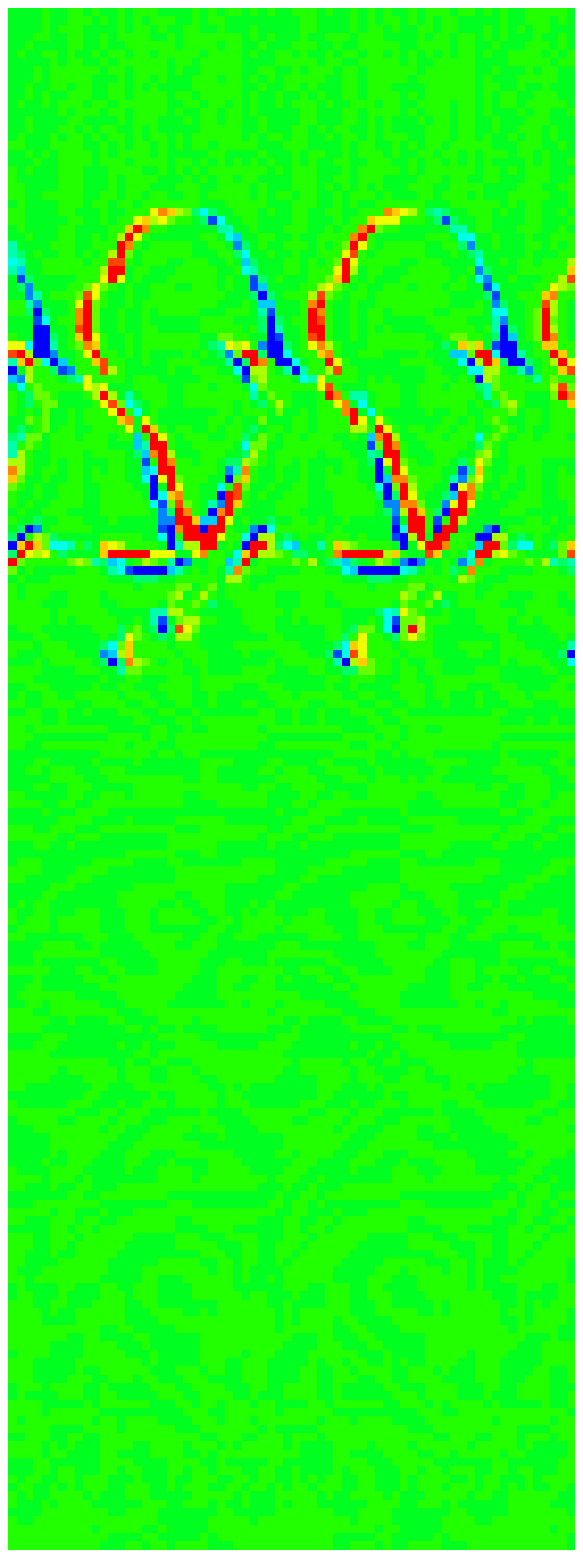}
\includegraphics[height=5.4cm,clip=true]{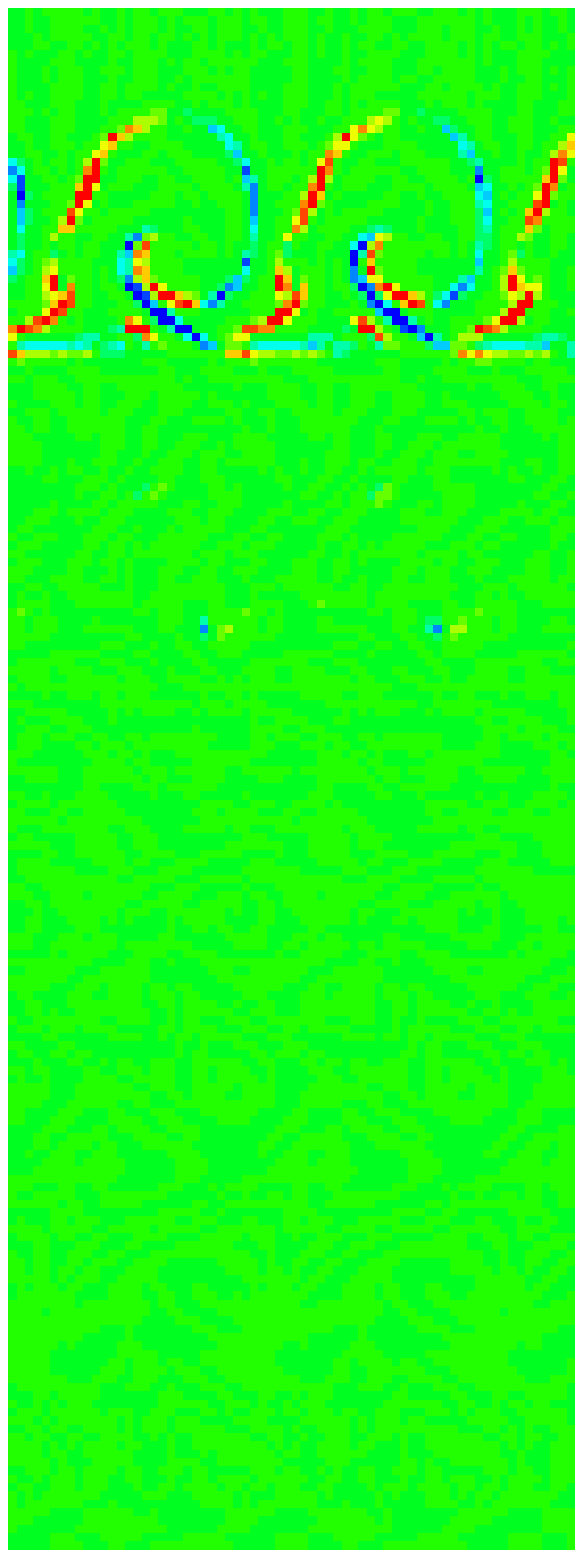}

\includegraphics[height=5.4cm,clip=true]{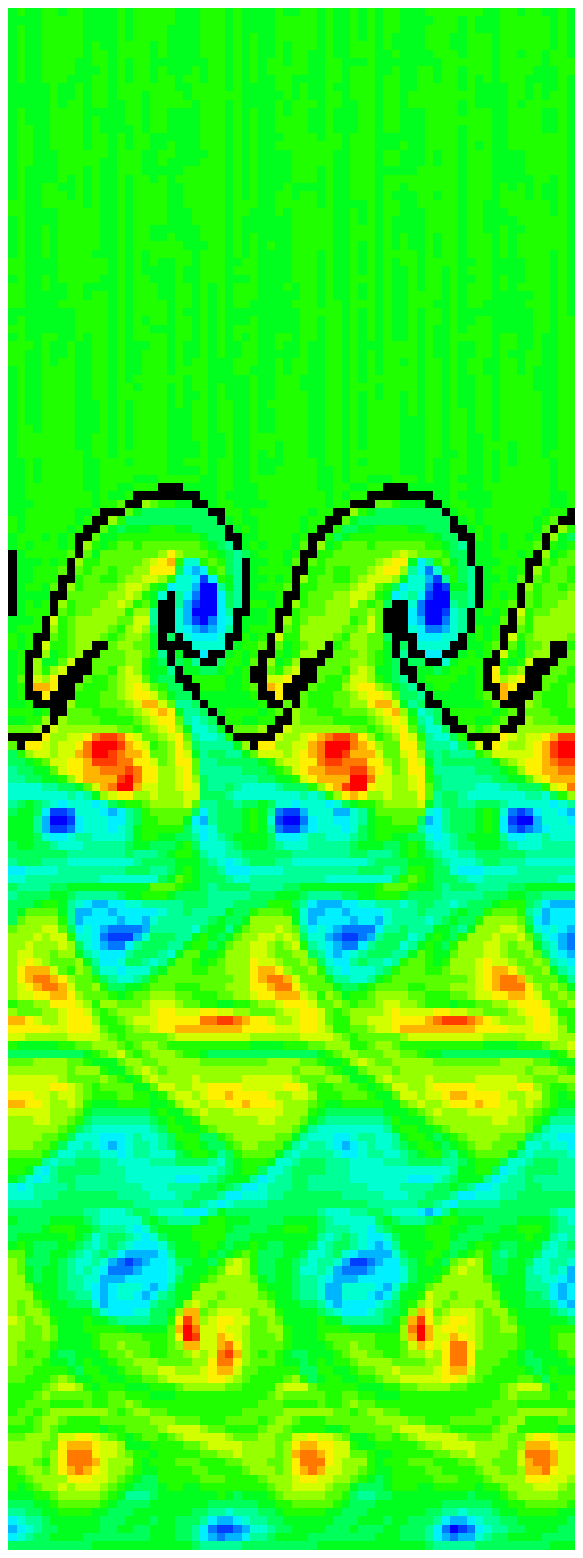}
\includegraphics[height=5.4cm,clip=true]{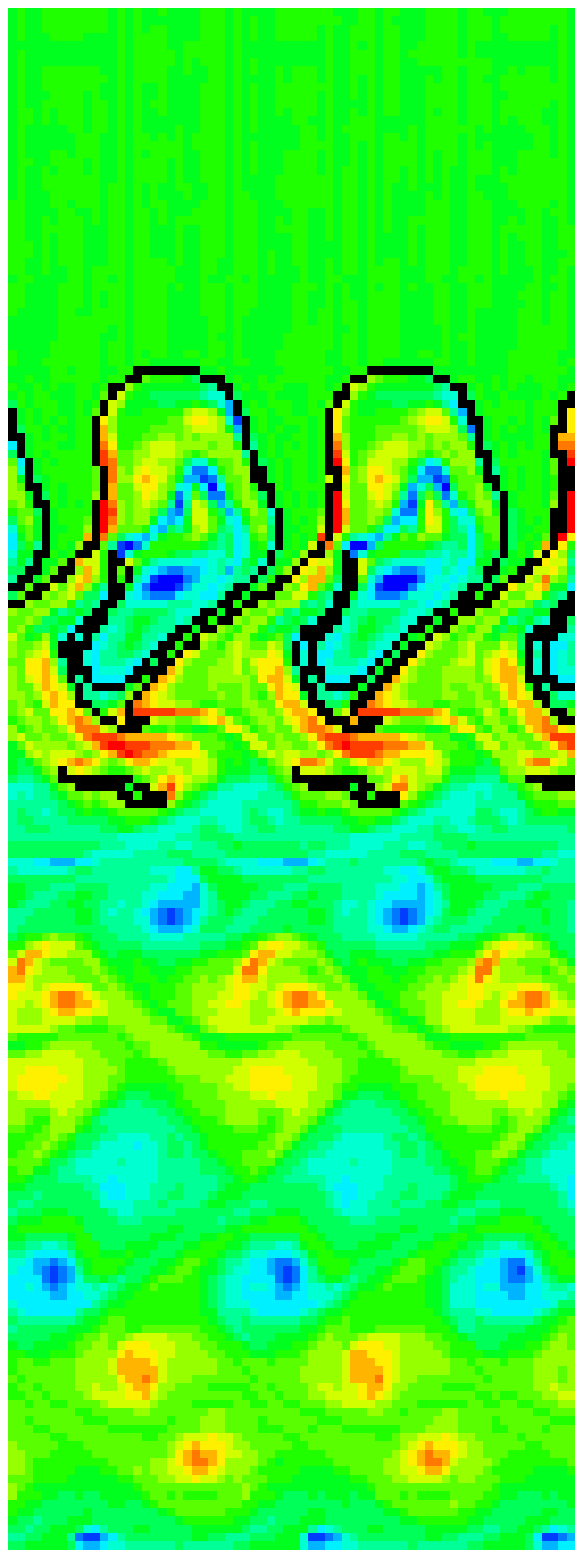}
\includegraphics[height=5.4cm,clip=true]{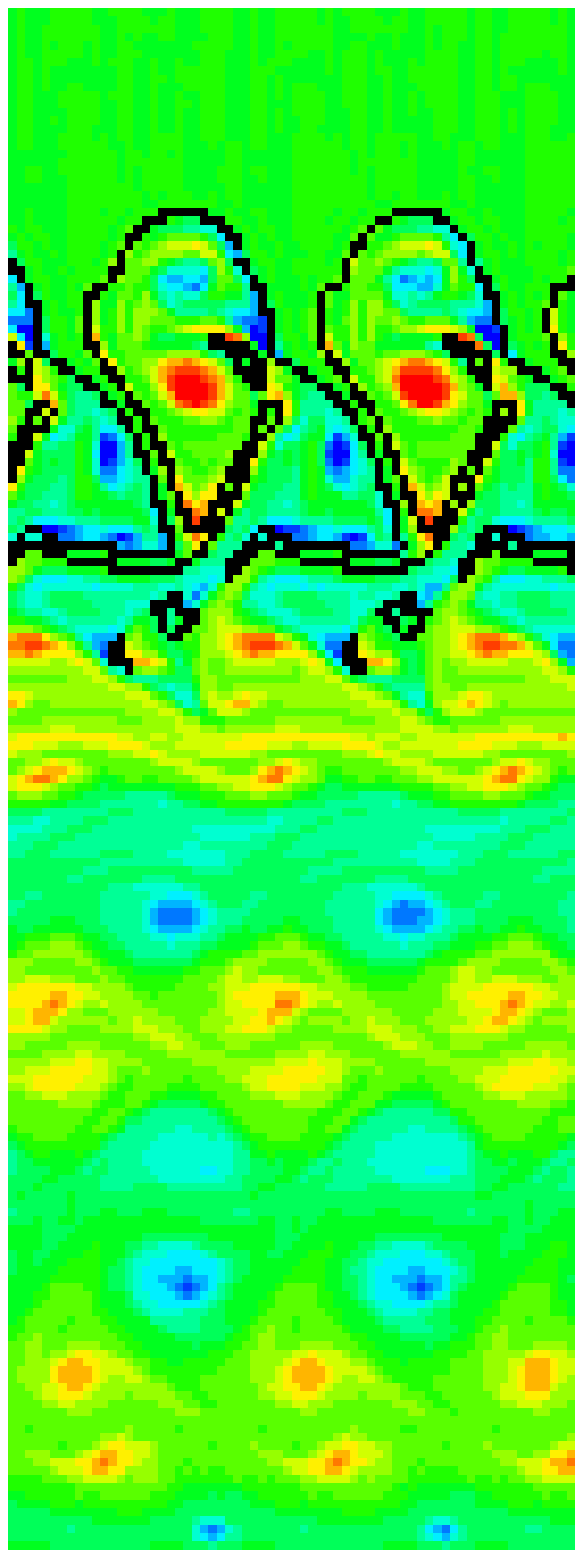}
\includegraphics[height=5.4cm,clip=true]{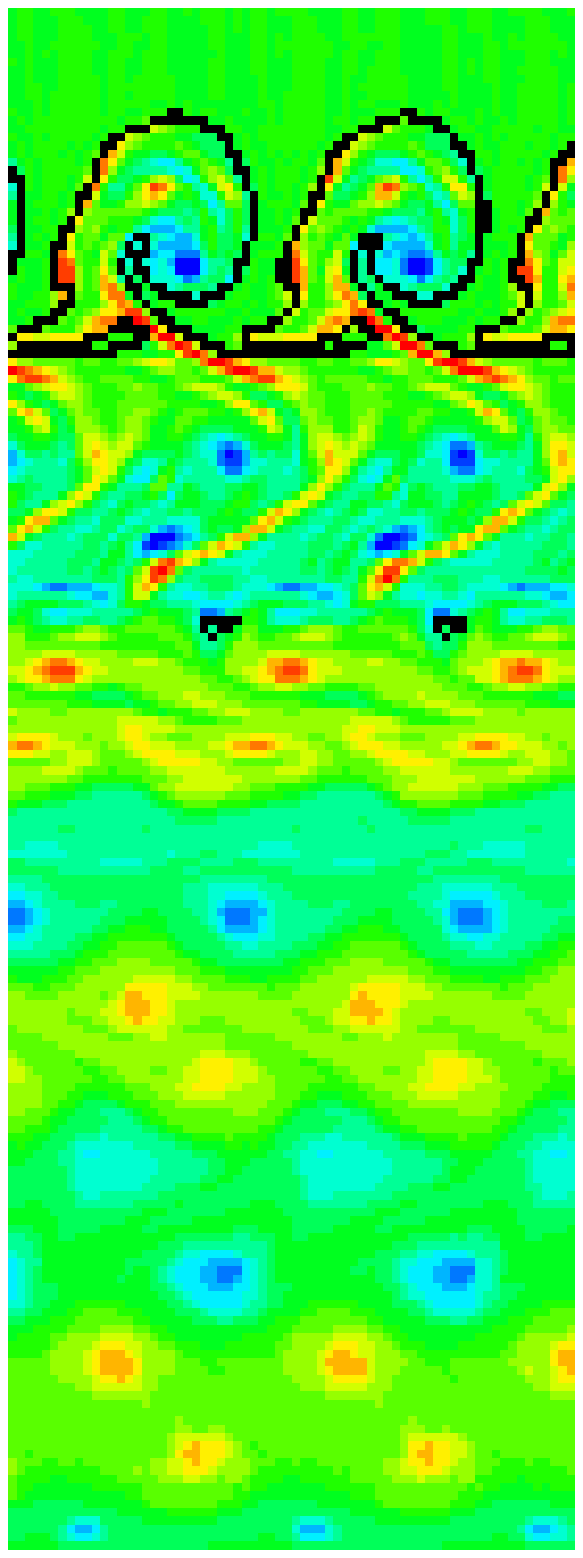}

\includegraphics[height=5.4cm,clip=true]{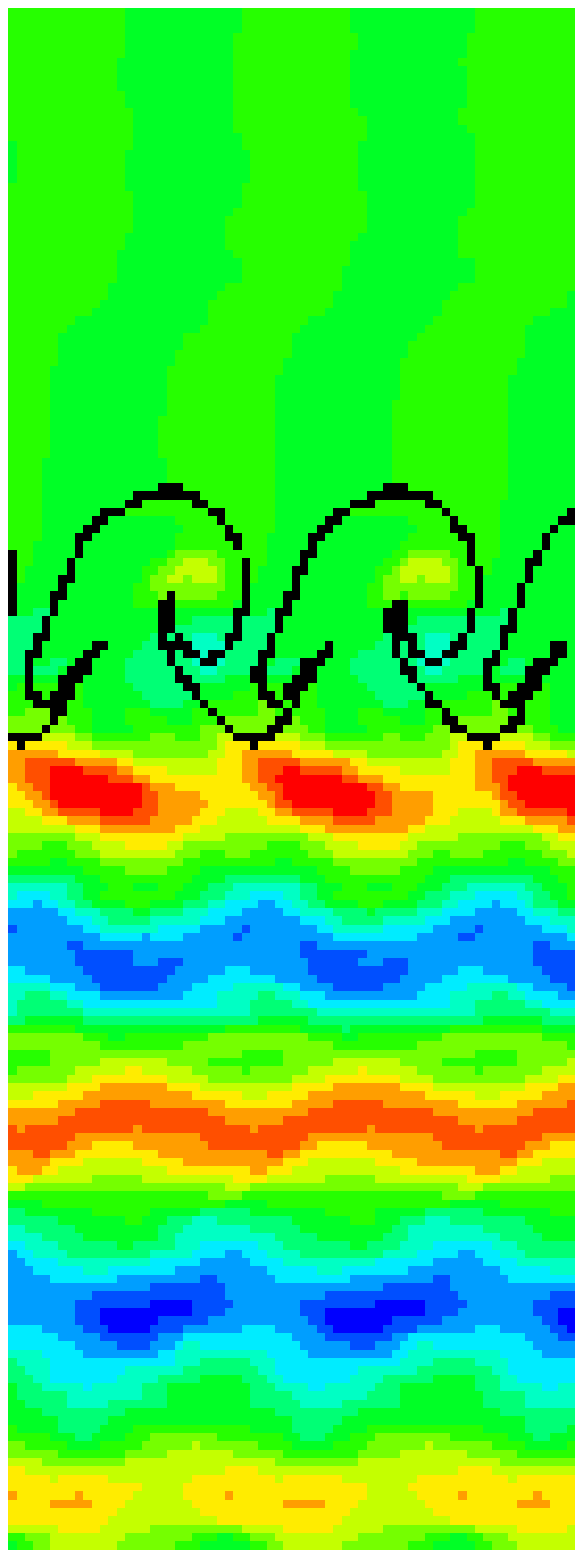}
\includegraphics[height=5.4cm,clip=true]{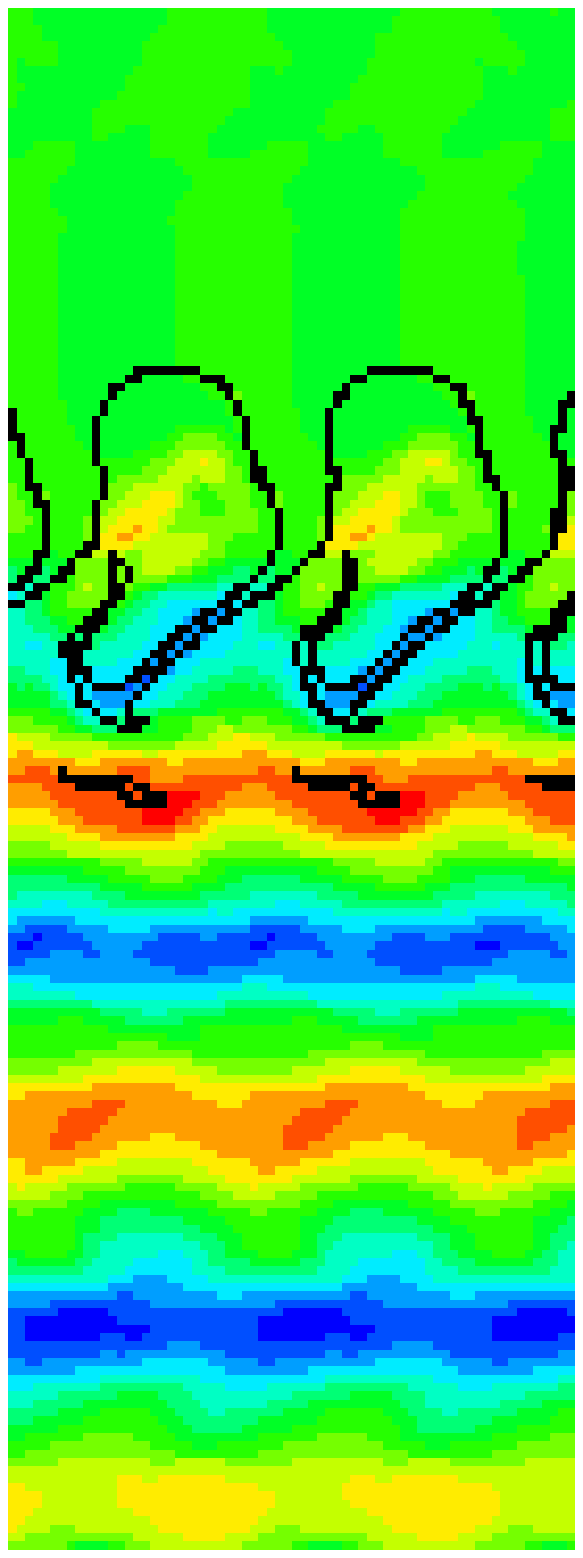}
\includegraphics[height=5.4cm,clip=true]{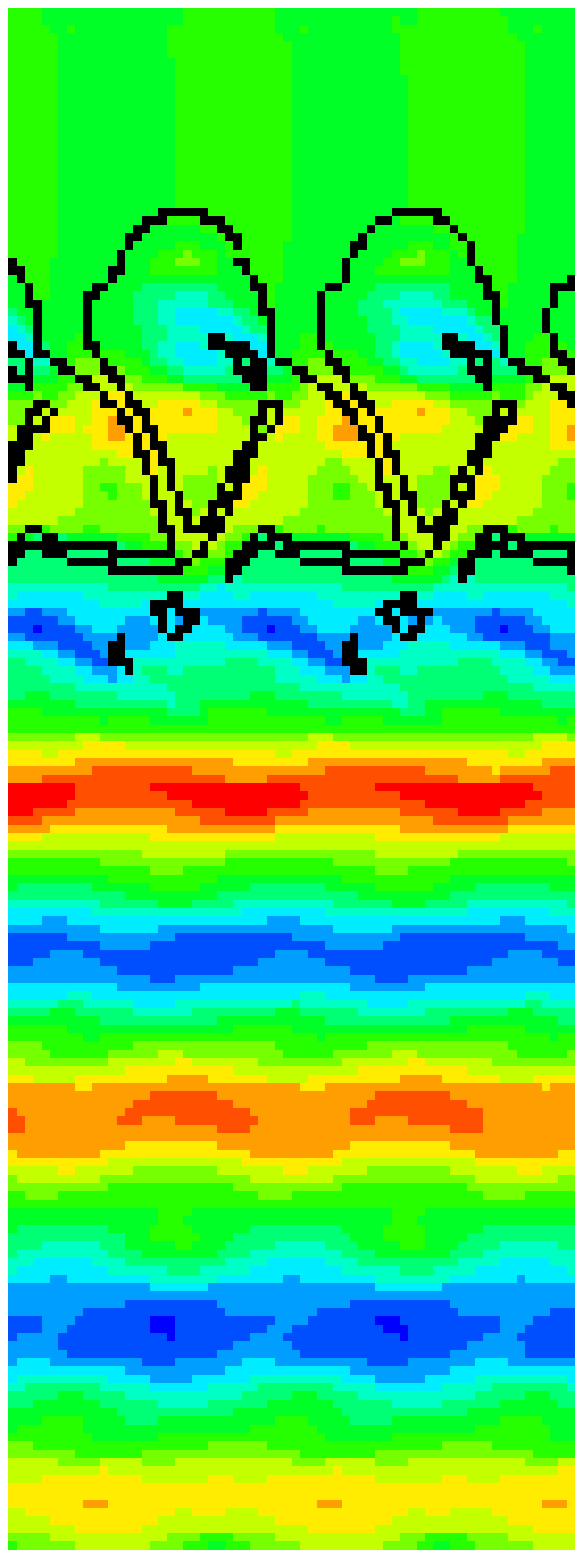}
\includegraphics[height=5.4cm,clip=true]{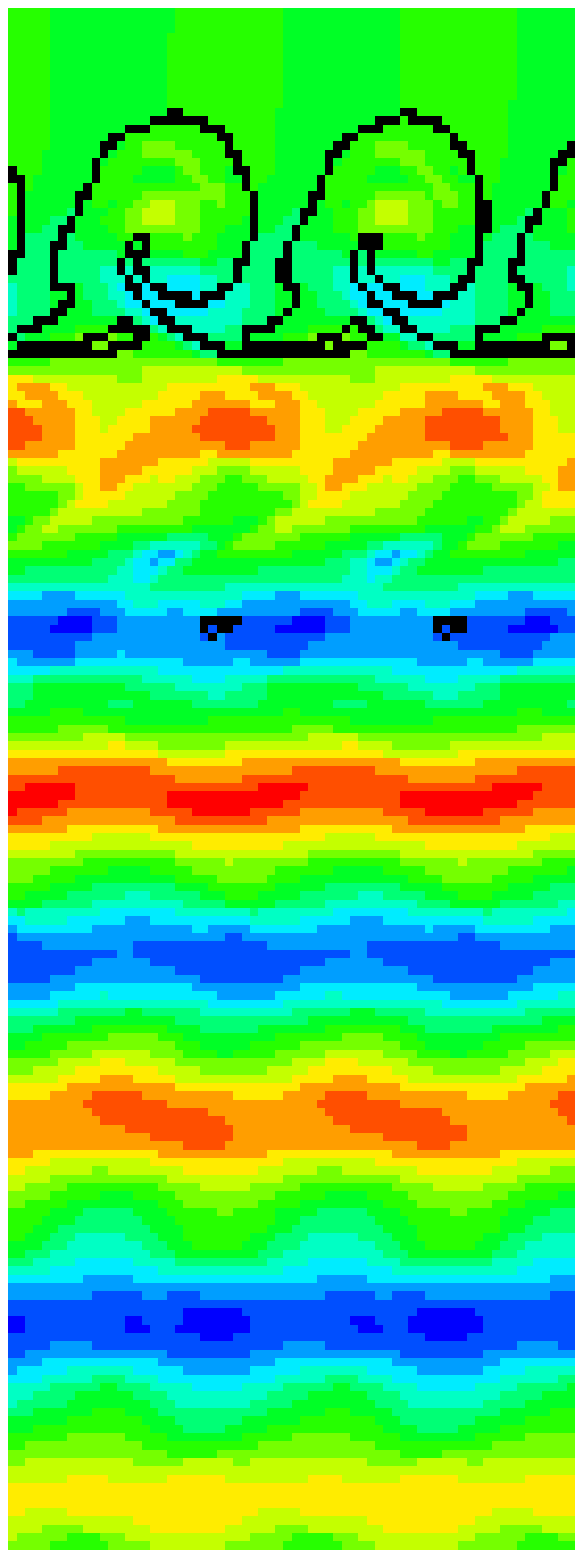}

\end{center}
\caption{Evolution of a turbulent flame in two-dimensions in statistical steady state. Shown is the
  distribution of vorticity generation, $\nabla \rho \times
  \nabla p$ (top row), vorticity (middle row) and horizontal velocity
  (bottom row) at four different 
  instants of time (from left to right: 0.41, 0.47, 0.54 and 0.58\, s, corresponding to
  ``A'' - ``D'' marked in the first panel of Fig.~\ref{f:Dtself}).
  Each row covers about two eddy turnover times ( $\sim$ 0.2\, s). Each panel is a lateral composite of  2.5 of our periodic domains, arranged
  for  visualization purposes.}
\label{f:2dflm}
\end{figure}

On the basis of these experiments, we conclude that flames subject to the Rayleigh-Taylor instability
are intrinsically oscillatory. In 2D, the rising bubbles filled with hot burned
material are analogous to a cylinder passing through a fluid.  It is well known that this schematic situation typically 
results in the formation of a vortex street in the wake of the cylinder \citep{white+03}.  Similarly, a vortex street also appears in the wake of
the  flame front in our simulations, and the periodicity of the structure is apparent in
Fig.~\ref{f:2dflm}.

\begin{figure}[h]
\begin{center}
\includegraphics[height=5.4cm,clip=true]{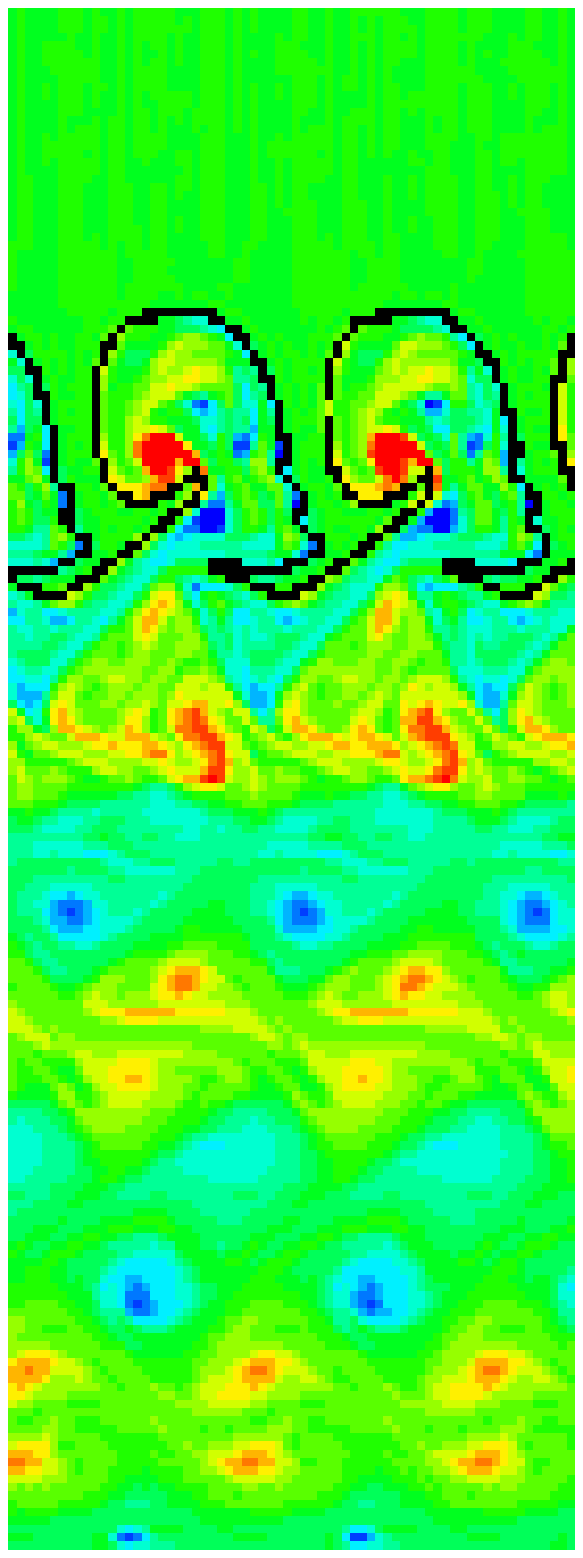}  
\includegraphics[height=5.4cm,clip=true]{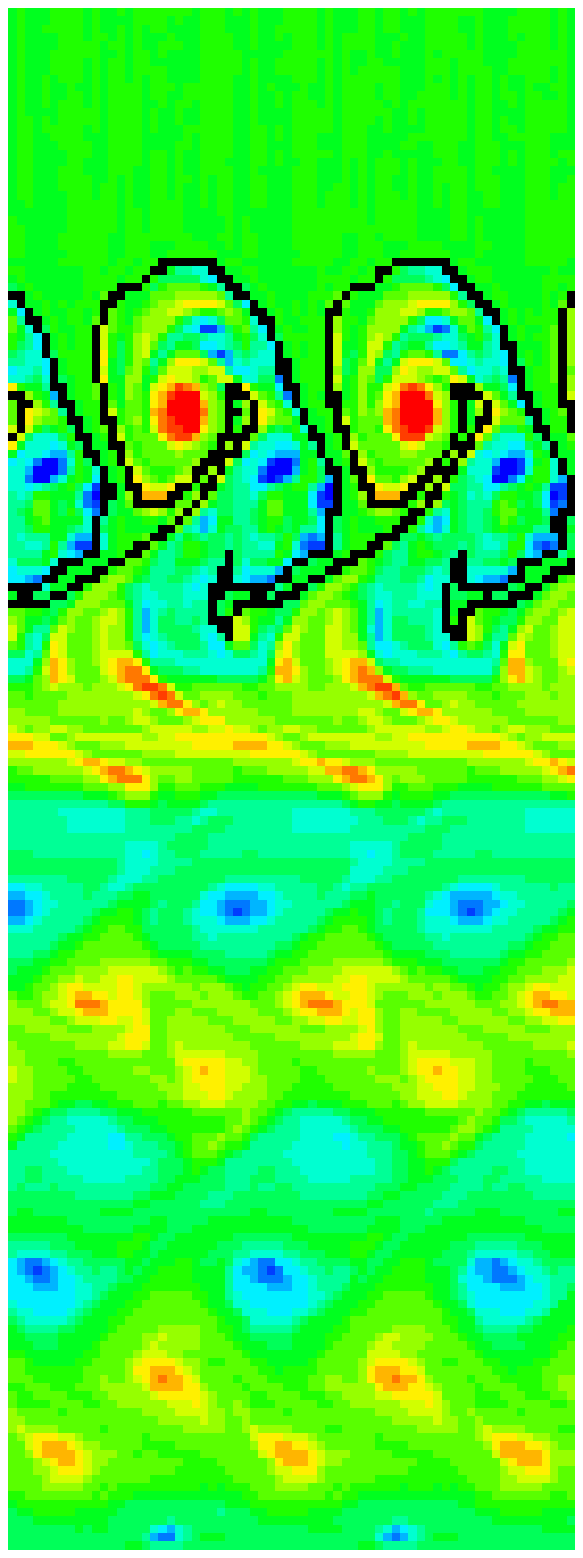}
\includegraphics[height=5.4cm,clip=true]{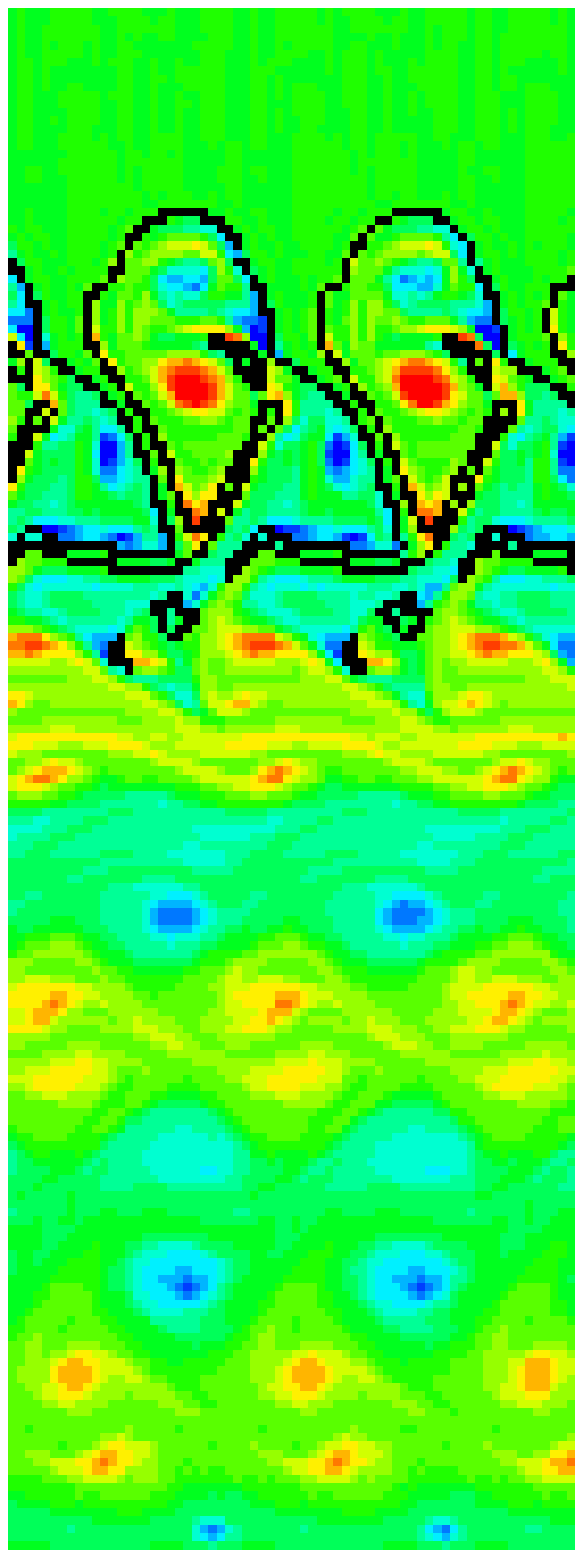}
\includegraphics[height=5.4cm,clip=true]{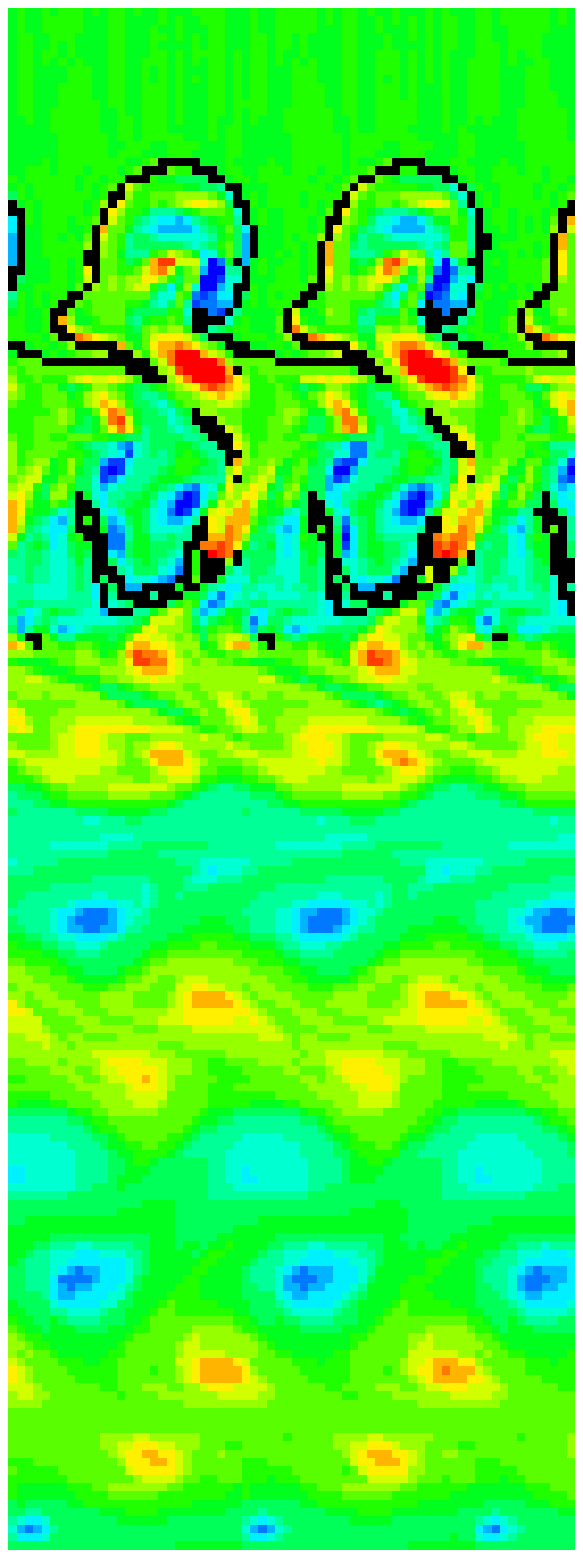}
\end{center}
\caption{Eddy turnover in a turbulent flame in two-dimensions. Plotted
is the vorticity distribution. Note the counter-clockwise rotating
eddy (the red ``blob''-like feature inside the bubble) as it stretches  and folds the
flame surface near the tail. Axes are the same as those in Fig.~\ref{f:2dflm}.}
\label{f:turnover}
\end{figure}

The production of vorticity (top panels in Fig.~\ref{f:2dflm})
is caused by the misalignment of the density and pressure gradients at
the sides of the flame bubbles. The pressure gradient tends to be
oriented along the direction of gravity, whereas the density
gradient is approximately normal to the flame surface. Vortices are
formed and then shed from the flame surface
(middle panels in Fig.~\ref{f:2dflm}). The vortices turn over, stretch
and fold the flame surface near the tail (panels for 0.47 and 0.54\, s).
These stretched and folded surfaces then collide and annihilate. During this
process, the heads of the bubbles wobble and their tails swing from
side to side. Approximately one full period of this  ``wobbling''
process is shown in Fig.~\ref{f:2dflm}. During each oscillation period, 
two eddies of opposite direction are shed from the surface of each bubble. One such turnover is illustrated in
Fig.~\ref{f:turnover}.

The size of the eddies is comparable to
the turbulent driving scale, $L$, {\it i.e.} the domain width.

Each flame wobble is characterized by
varying flame surface resulting in variations of the turbulent flame speed.
For example, consider two maxima of the turbulent flame speed marked ``B'' and ``C'' in the top panel
of Fig.~\ref{f:Dtself} (2D model with $D_l = 1.07e6$ cm/s). The flame structure corresponding to these two extrema is
illustrated in the second and the third column in Fig.~\ref{f:2dflm}. Note that,
at these two instants of time, the flame surface area is also at its largest, as
indicated by extreme length of the bubble's tail.

The above picture is in qualitative agreement with the recent
two-dimensional study of \citet{vladimirova+05b} obtained in the
Boussinesq limit. In particular, the morphology of the flame and the formation
of vortex streets are common features of the two models
[see \citep{vladimirova+05b}, panels (a),
  (e), and (f) in Fig.~1].  Note that in their
simulations with symmetric boundaries, no oscillatory behavior is
observed. We consider this to 
be an effect of the symmetric boundary conditions. In reality, there is no imposed
symmetry and the RT flame is physically periodic as a result of eddy
turnover. It should be pointed out the horizontal reflecting boundary
they sometimes term a symmetric boundary should not be confused with a
wall boundary. Here, we observe periodic
behavior also in the flame evolution with horizontal wall boundaries. In
addition, \citet{vladimirova+05b} surprisingly find that the ratio of 
the vertical wavelength and the domain width is invariant and always
close to 2.  A similar result is  
obtained here (Fig.~\ref{f:flm_2d_Dl}) for which we offer the following explanation:

\begin{figure}[h]
\begin{center}
\includegraphics[height=5.5cm,clip=true]{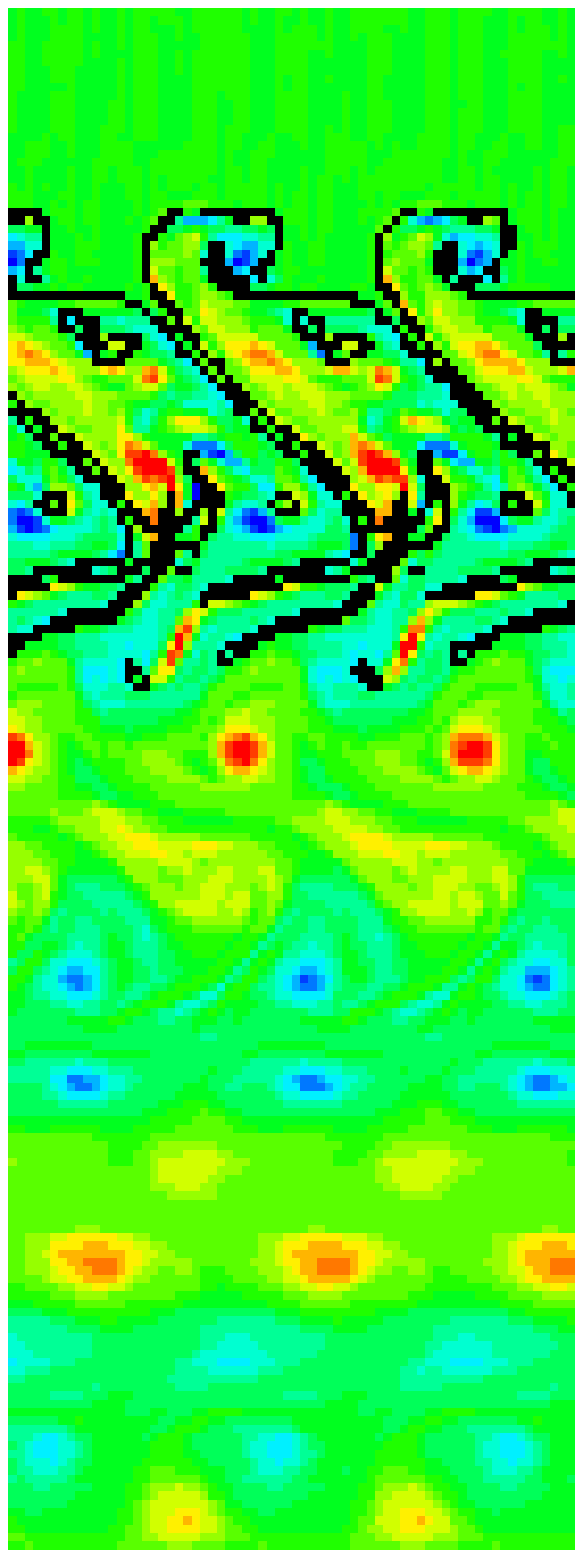}
\includegraphics[height=5.5cm,clip=true]{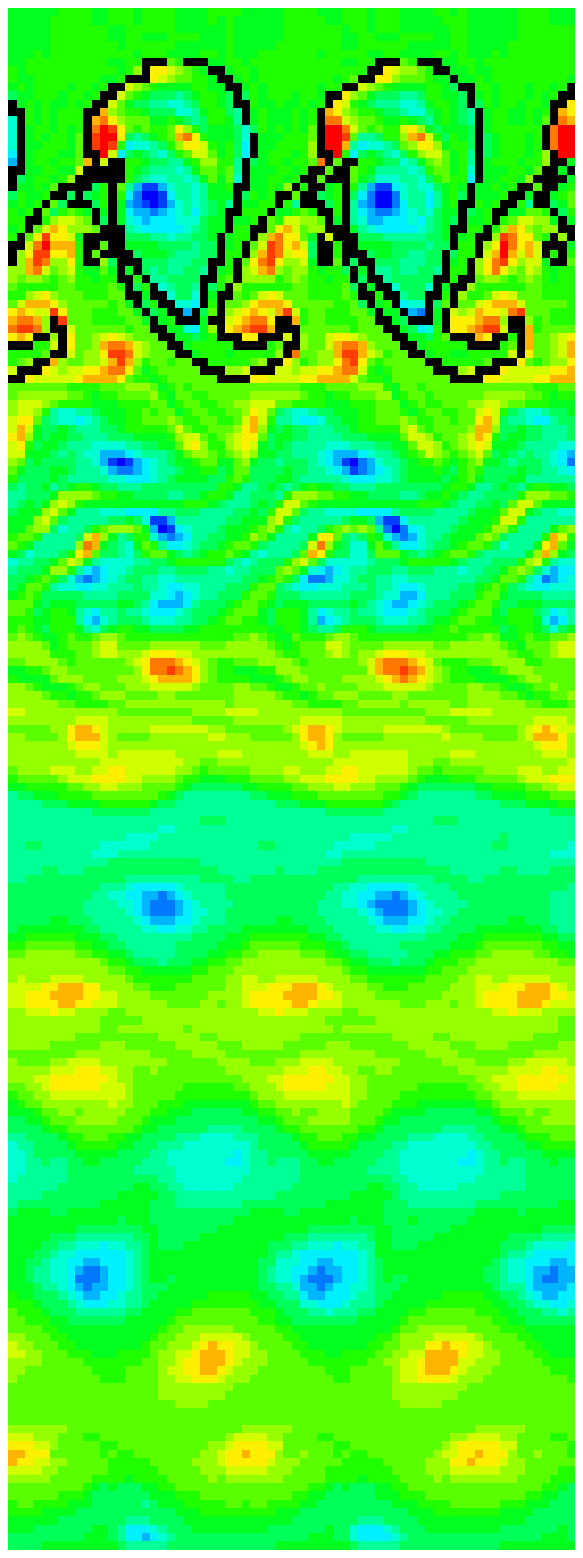}
\includegraphics[height=5.5cm,clip=true]{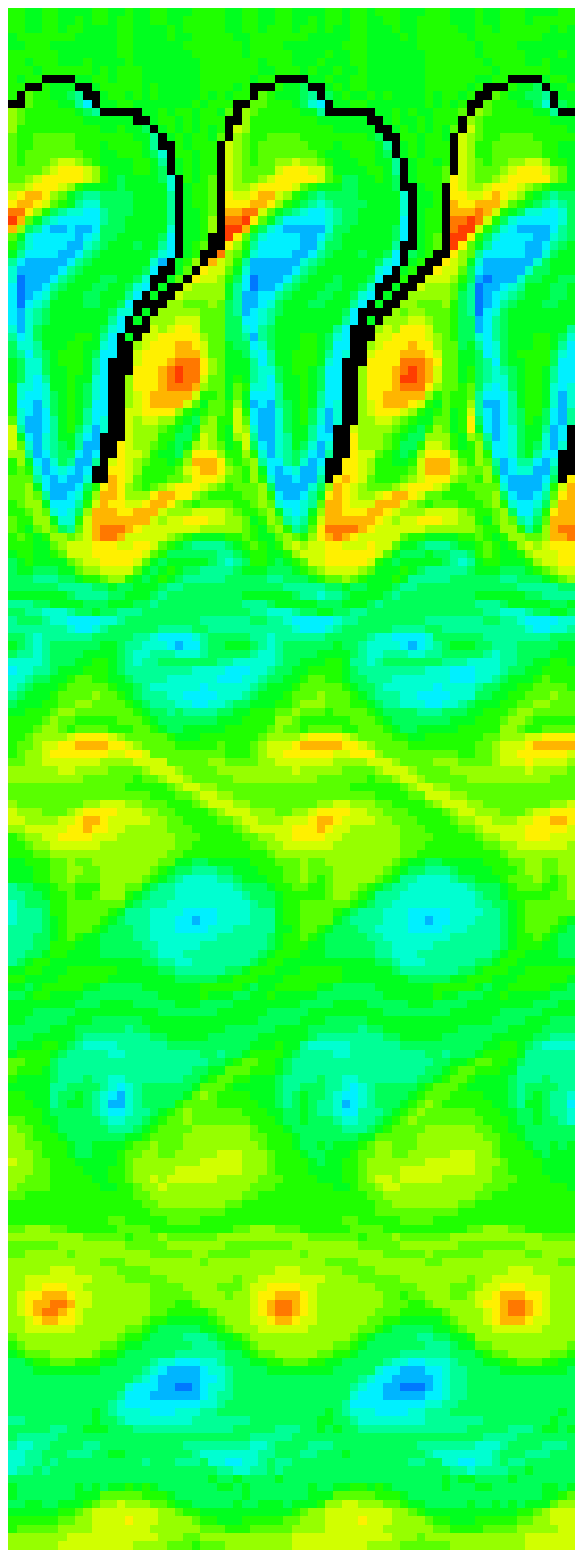}
\end{center}
\caption{Vorticity distributions in steady state. Laminar flame speeds
  are (from left to right): 0.54e6 cm/s, 1.07e6 cm/s and 2.14e6 cm/s
  corresponding to the three 2D runs shown in Fig.~\ref{f:S_Dl}. The scale
  of the axes are kept the same in each panel. The vertical wavelengths are
  approximately the same, being about twice as large as the domain
  width. Note, again, each panel has been duplicated horizontally 2.5
  times and axes are the same as those in Fig.~\ref{f:2dflm}.}
\label{f:flm_2d_Dl}
\end{figure}

As mentioned above, in steady state, the largest RT unstable structure
is always comparable to the turbulent driving scale of  the
system. Since the periodicity is caused by the turnover of these
largest eddies, and there are two turnovers in each period, it is 
natural that the vertical wavelength is always about twice as large as
the domain width. 

\begin{figure}[ht]
\begin{center}
\includegraphics[height=5.5cm,clip=true]{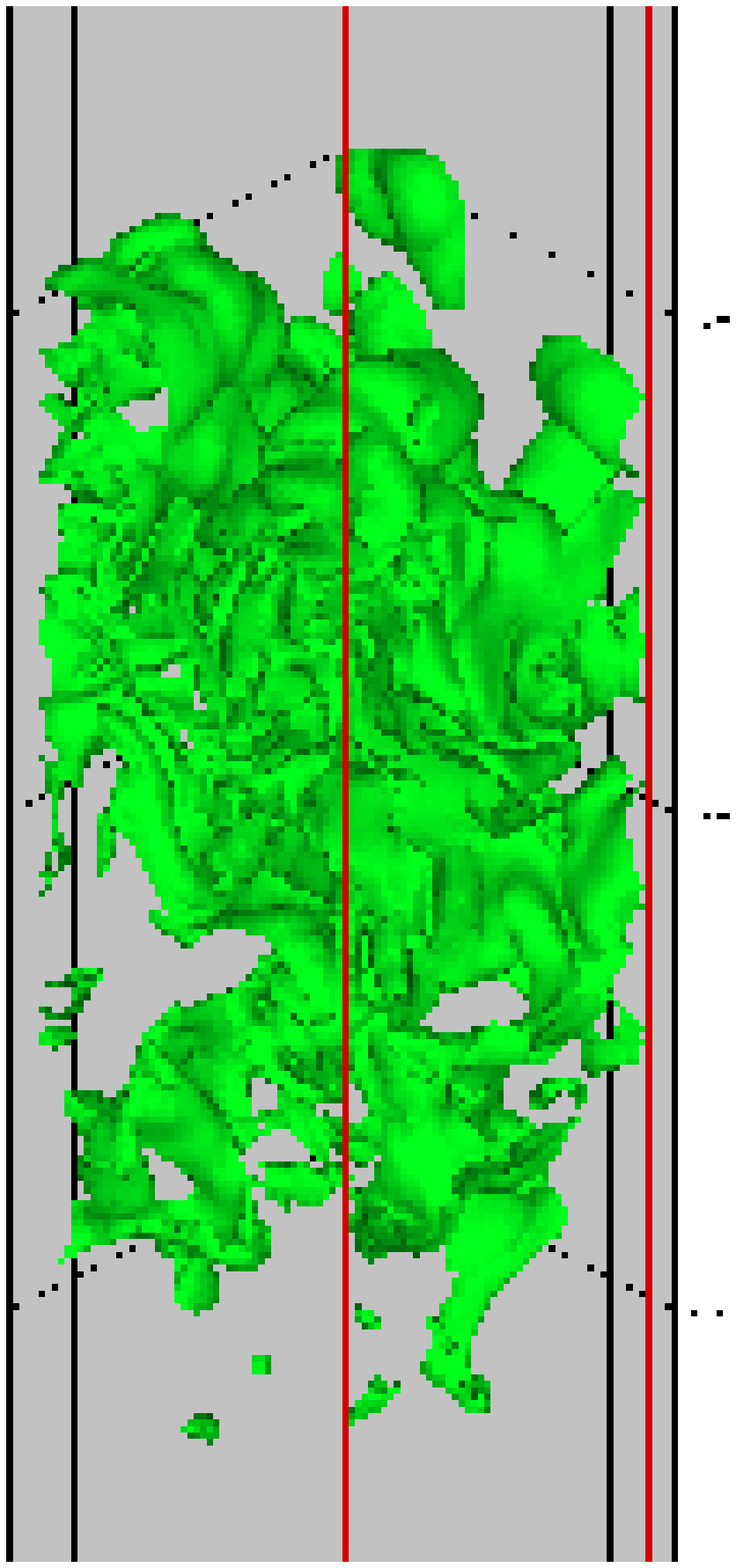}
\includegraphics[height=5.5cm,clip=true]{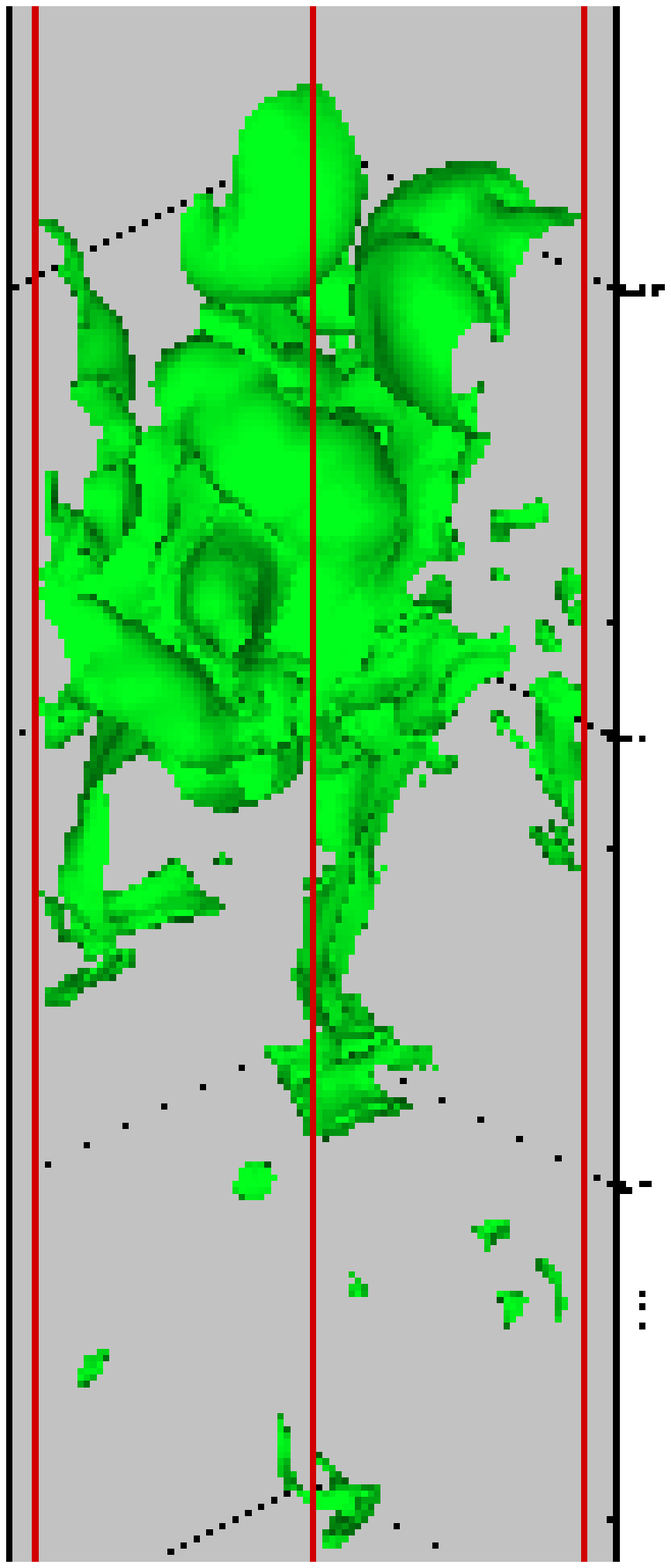}
\includegraphics[height=5.5cm,clip=true]{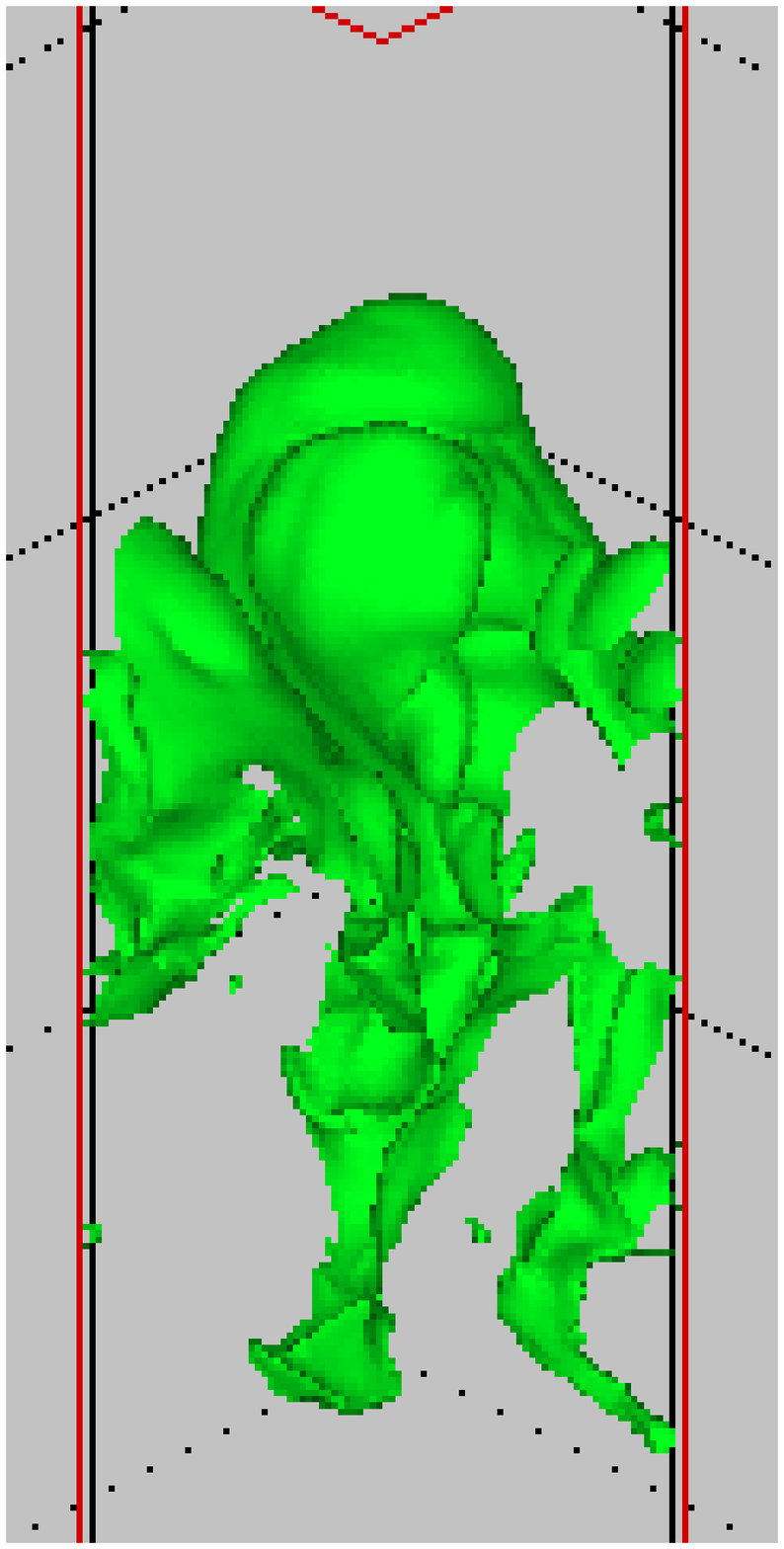}
\end{center}
\caption{Morphology of the turbulent flame surface in models with
  different laminar flame speeds at late times. (left):
  model DH, $D_l = 0.54 \times 10^6$ cm/s; (middle): model R1, $D_l = 1.07
  \times 10^6$ cm/s; (right) model D2, $D_l = 2.14 \times 10^6$
  cm/s. Note the larger the laminar flame speed, the smoother and, concomitantly,  the
  smaller the flame surface. The vertical spatial scale are $1 \times 10^7$ to $1.6 \times 10^7$, $0.7 \times 10^7$ to $1.35 \times 10^7$ and $1.2 \times 10^7$ to $1.8 \times 10^7$ cm for model DH, R1 and D2, respectively.}
\label{f:flm_Dl_3d}
\end{figure}

The character of the steady state has been discussed in the context of
2D simulations in the above. In 3D, although vortices shed from the
flame surface do not merge into the bulk but cascade into smaller and
smaller structures, and the flame surface is not as regular
(Fig.~\ref{f:flm_Dl_3d}),

%
%
%

the overall mechanism of vorticity production near the flame front and
the subsequent interaction between vortices and the flame surface is
similar to that observed in 2D. Regarding vorticity generation, vortex
stretching is an important contributing mechanism of vorticty
generation uniquely present in 3D. While vortex stretching is a
dominant source of vorticity in the bulk, its strength is comparable
to that of barcolinic vorticity generation near the flame surface. The
latter mechanism is common to 2D and 3D and is the original source of
vorticity in our problem.

\begin{figure}[h]
\begin{center}
\includegraphics[height=5.5cm,clip=true]{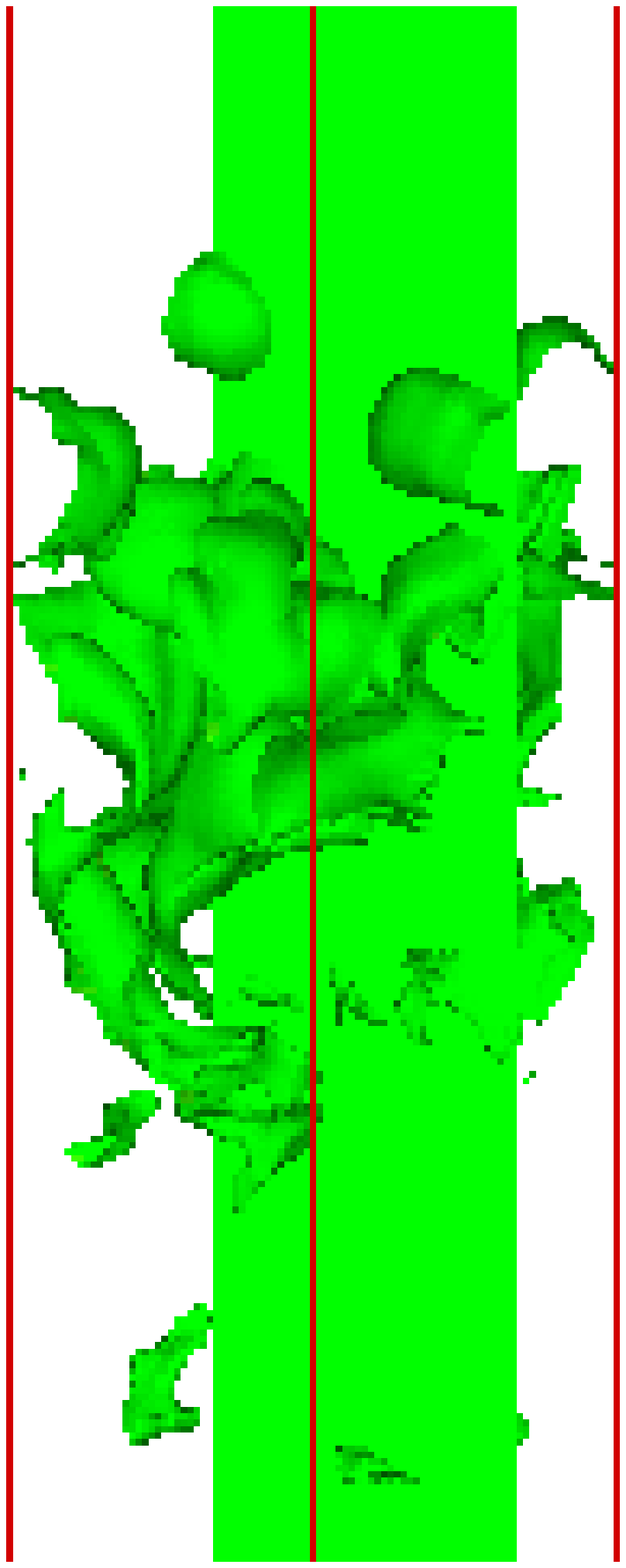} 
\includegraphics[height=6.5cm,clip=true]{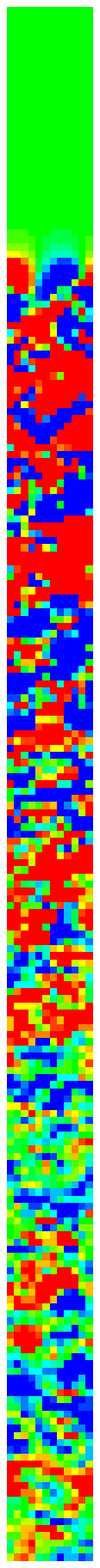}
\includegraphics[height=6.5cm,clip=true]{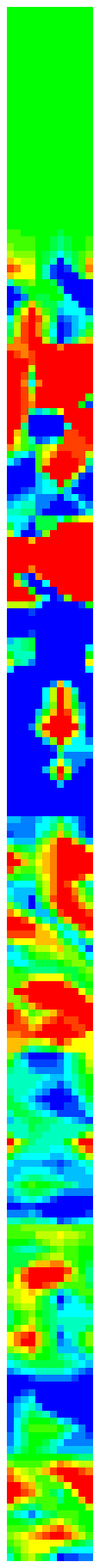}
\includegraphics[height=6.5cm,clip=true]{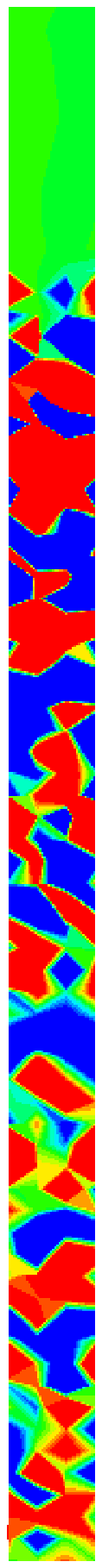}
\end{center}
\caption{Turbulent flame at $\sim$0.83 s for model R1. The 3D flame surface
  and the position of the 2D slice are shown in the first panel. Only
  that part of the domain that contains the flame (vertical spatial scale from $1 \times 10^7$ to $1.6 \times 10^7$ cm) is shown in this first  panel. The second
  panel plots the horizontal velocity field (component in the plane)
  in the 2D slice. The 3rd and 4th panel show the horizontal velocity
  field and the vorticity (normal to the plane) on the coarsened
  grid. Note the trace of the layered structure in these two
  panels. (N.B. In contrast to the 3D plot, the slices cover the entire domain height from 0 to $3 \times 10^7$ cm.)}
\label{f:Layer_3d}
\end{figure}

In addition, the same semi-periodic behavior is also observed
in 3D flame evolution, as can be seen in Fig.~\ref{f:Dtself}. Just as
in 2D, this semi-periodic behavior is caused by the turnover of
largest eddies. Despite the complicated structure of a 3D flame, the
periodically layered structure can still be discerned in a 2D slice of
the 3D flow by coarsening the grid by a factor of 16. Such slices are
shown in Fig.~\ref{f:Layer_3d}.

\begin{figure}[ht]
\begin{center}
\includegraphics[height=5.5cm,clip=true]{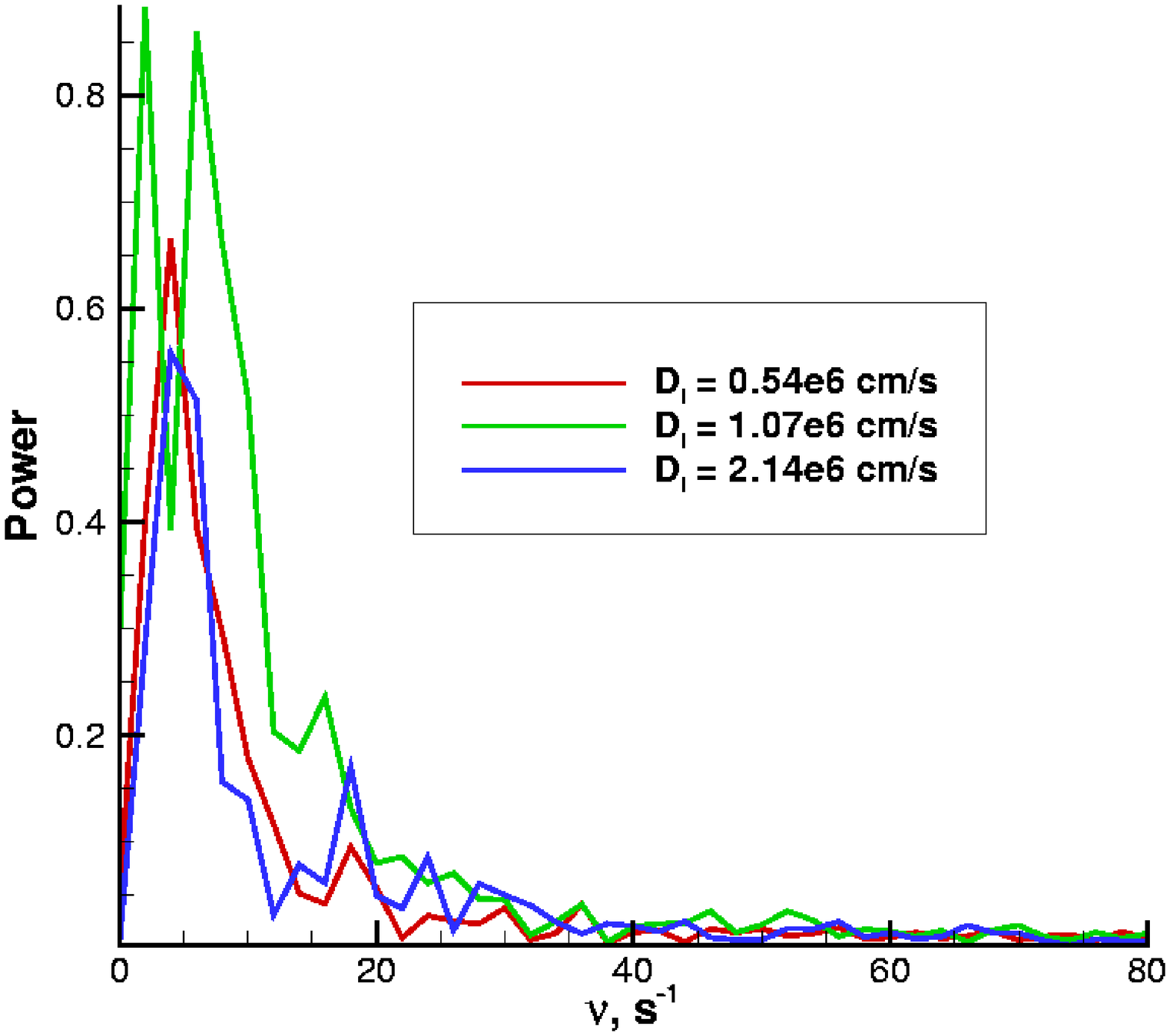}
\includegraphics[height=5.5cm,clip=true]{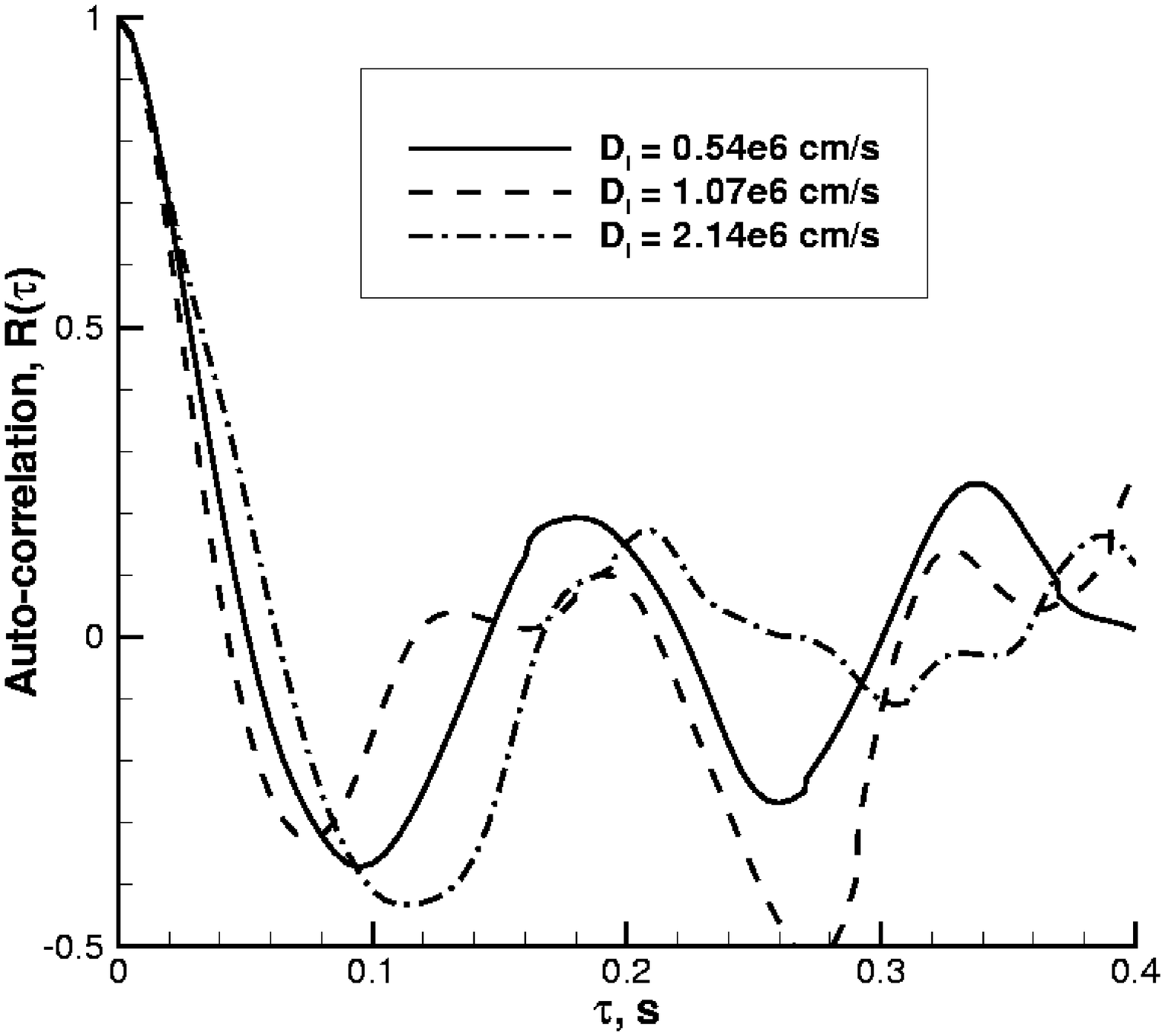}
\end{center}
\caption{Time scales for the evolution of turbulent flames.  The {\sc
clean} power spectrum (left) and the auto-correlation function $R(\tau)$ (right)
are plotted for the 3D time series data shown in Fig.~\ref{f:Dtself}. In the first panel, the
power peaks at a frequency of $\approx 5$ $s^{-1}$. This is consistent
with the separation of the first and the second peak of the
auto-correlation by $\approx 0.2$\,s.}
\label{f:t_power}
\end{figure}

To actually extract the period of the temporal evolution of the flame,
the run shown in (Fig.~\ref{f:Dtself}) was analyzed for $t \ge 0.6$\,s
in both 2D and 3D. Figure~\ref{f:t_power} shows the results of the
analysis obtained for 3D with the help of the {\sc clean} method
\citep{roberts+87} for power spectrum analysis, together with the
results obtained by computing the auto-correlation of the time series
$D_t$ (at $t \ge 0.6$\,s) in Fig.~\ref{f:Dtself} , {\it i.e.} the
correlation of $D_t(t)$ against a time-shifted (by $\tau$) version of
it. Fig.~\ref{f:t_power}
%
%
%

\begin{figure}[ht]
\begin{center}
\includegraphics[height=7.cm,clip=true]{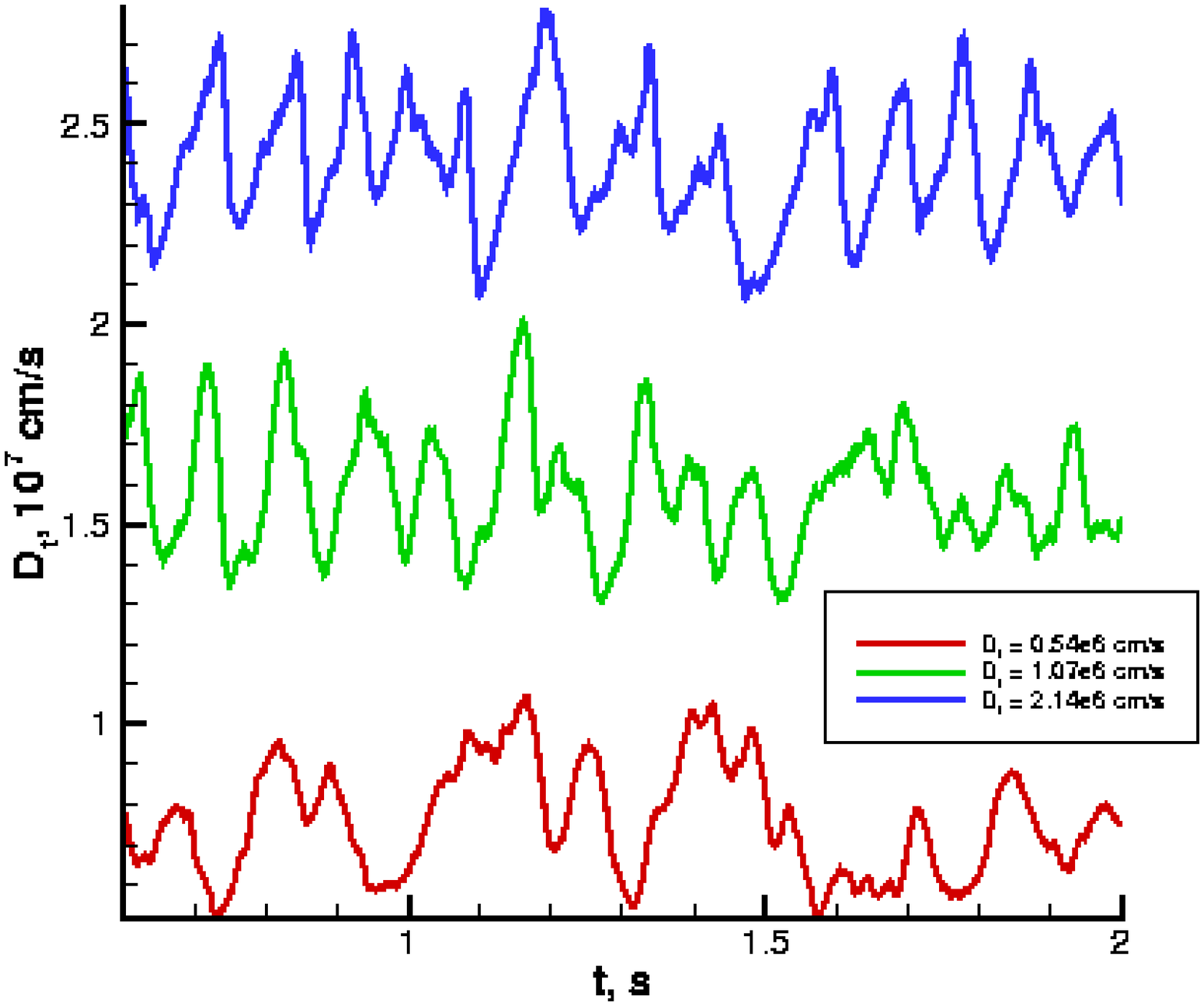}
\includegraphics[height=5.5cm,clip=true]{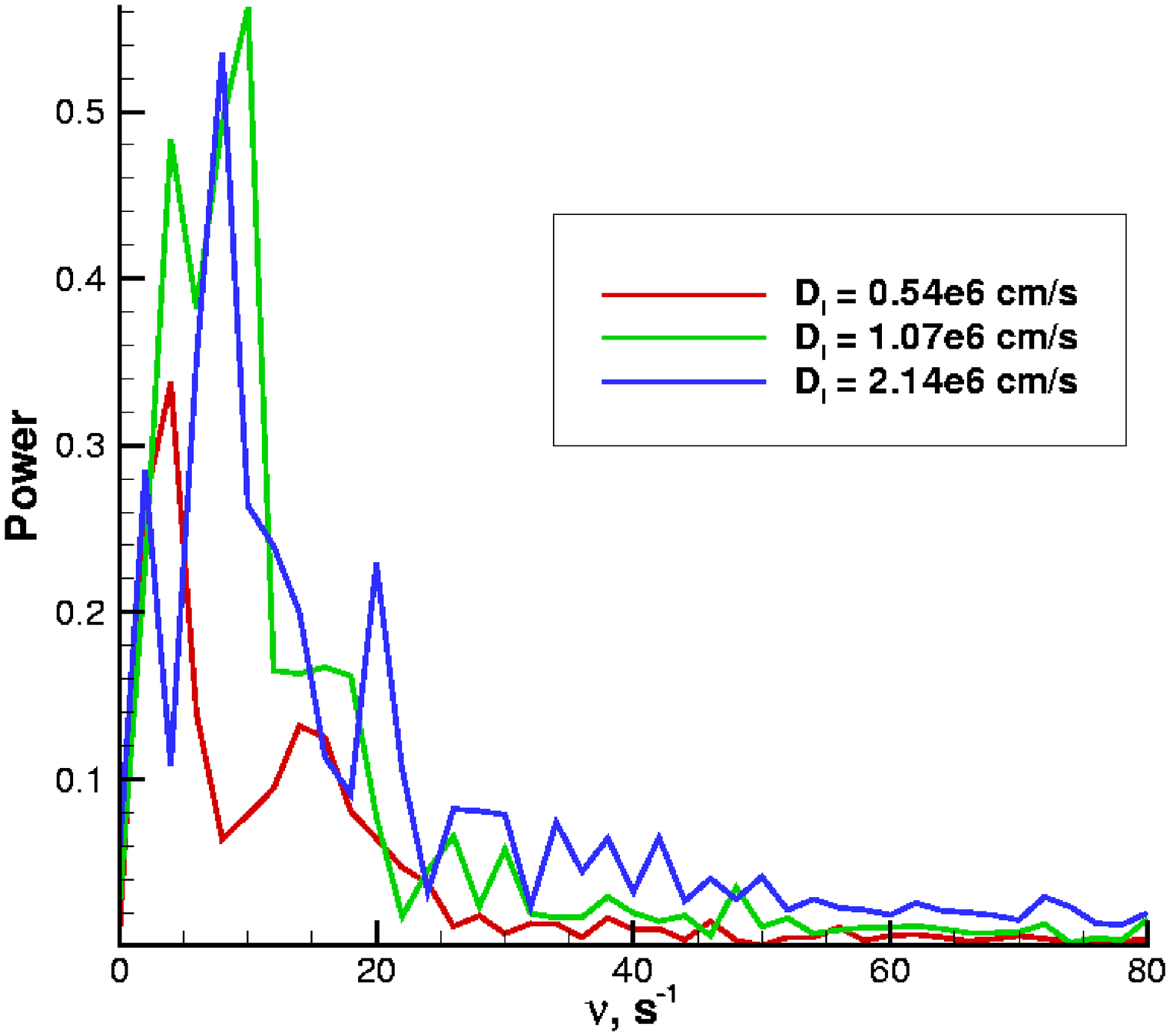}
\end{center}
\caption{Evolution of the turbulent flame speed, $D_t$ in 2D in steady state (left panel) 
     and the corresponding CLEAN power spectrum (right panel). Note that offsets of $0.9 \times 10^7$ and $1.8 \times 10^7$ cm/s have been added to the slowest and the second slowest laminar flame speed cases, respectively.}
\label{f:t_power_2d}
\end{figure}

shows that both methods indicate the presence of excess power around $ 0.2
s$. In particular, there is a clear accumulation of power for
frequencies $< 10$ s$^{-1}$ in the power spectrum calculation. The
first and second peaks of the auto-correlation function are separated
by $\sim 0.2$\,s, in excellent agreement with presence of power at a
frequency $\approx 5$ $s^{-1}$ in the {\sc clean} result.
Fig.~\ref{f:t_power_2d} 

shows the corresponding {\sc clean} results for three selected 2D
models. Since, as discussed in Sec.~\ref{s:S_evol}, there are two
eddy turnovers during one period, we expect the power to be 
concentrated at low frequencies corresponding to one (half period) and two
(full period) eddy turnover times.  This expectation is confirmed, as can be seen in
Fig.~\ref{f:t_power_2d}.

This  period determination is consistent with the time for the flame
``brush'' to travel the distance between the vortex streets
left in the bubble wake. In the particular case shown in Fig.~\ref{f:2dflm}, the distance between two consecutive, countervailing
vortex streets is $\sim 2.7 \times 10^6$~cm (this can also be
measured in the bottom panels of Fig.~\ref{f:2dflm} more easily), while the flame sweeps through the
computational domain with total speed $\sim 1.5
\times 10^7$ cm/s.\footnote{The total speed is the sum of
the turbulent flame speed, $\sim 0.8 \times 10^6$ cm/s (Fig.~\ref{f:Dtself}), and the mean
fluid velocity, $\sim 0.7 \times 10^6$ cm/s.} 

Our results in this section lead to the conclusion that the semi-periodic
behavior we observe is caused by the turnover of the 
largest eddies present in the simulation. This behavior is clearly
visible in 2D simulations.
In the supernova setting, a steady state realized in our model
flame study is, strictly speaking, not present due to supernova
expansion that makes both gravity and pre-flame conditions
time-dependent. Therefore, supernova flame never achieves steady state
but rather constantly struggles to approach that (now) time-dependent
condition. One may expect that in that case a semi-periodic behavior
exhibited by model flames might be significantly weaker or perhaps
even completely suppressed (depending on whether flame evolution
proceeds on timescales short compared to that of the background
expansion).
\subsection{Governing Equation for Flame Surface Area Evolution}\label{s:S2}
As previously mentioned, \cite{khokhlov95} suggested that the flame surface evolution is
governed by Eq.~(\ref{e:dSdt}). The first term on the right-hand
side of this equation describes flame surface creation via strain, while the
second term describes flame surface destruction due to the
propagation of cusps. \citet{khokhlov95} motivated a general $S^2$
dependence of the flame surface destruction term by considering elemental
geometries such as intersecting spheres and parallel planes. 

To determine whether the flame surface destruction term in
(\ref{e:dSdt}) is indeed proportional to  S$^2$ let us consider
the following generalized equation:

\begin{equation} \label{e:dSdt_Sn}
  \frac{dS}{dt} = aS - bS^n\,,
\end{equation}
where $a = c$ and $b = dD_l$ are arbitary constants for the moment, and $n >$
0. A steady state (equilibrium) solution, $S_e$, to (\ref{e:dSdt_Sn})
can be found by equating the creation (first) and destruction
(second) term to produce,

\begin{equation} \label{e:Se}
  S_e = \left(\frac{a}{b}\right)^\frac{1}{n-1}.
\end{equation}
Now, let us add a constant perturbation $\Delta a$ to the surface
creation coefficient, $a$, to obtain a
new equilibrium solution $S_e^\prime$ as,

\begin{equation} \label{e:Sen}
  S_e^\prime = \left(\frac{a + \Delta a}{b}\right)^\frac{1}{n-1} =
  \left(\frac{a}{b}\right)^\frac{1}{n-1}\left(1 + \frac{\Delta a}{a}\right)^\frac{1}{n-1}\,,
\end{equation}
Therefore, the difference between the two equilibrium solutions
is

\begin{equation} \label{e:delSe}
\begin{array}{ll} 
  \Delta S &= S_e^\prime - S_e \\ &= \left(\frac{a + \Delta a}{b}\right)^\frac{1}{n-1}
  - \left(\frac{a}{b}\right)^\frac{1}{n-1} \\ & =
  \left(\frac{a}{b}\right)^\frac{1}{n-1}[\left(1 + \frac{\Delta
  a}{a}\right)^\frac{1}{n-1}-1].
\end{array}  
\end{equation}
In the limit of $\frac{\Delta a}{a} \ll 1$, Eq.~(\ref{e:delSe}) yields

\begin{equation} \label{e:delSelim}
  \Delta S  = 
  \frac{\Delta a}{(n-1)a}\left(\frac{a}{b}\right)^\frac{1}{n-1}.
\end{equation}
>From this one can see that for a given $a$, $b$ and $n > 1$, the magnitude of
$\Delta S$ is proportional to $\Delta a$. Also, 
for a given $a$, $\Delta a$ and $n > 1$, $\Delta S$ is proportional to
$\left(\frac{1}{b}\right)^\frac{1}{n-1}$. 

\begin{figure}[ht]
\begin{center}
\includegraphics[height=7.cm,clip=true]{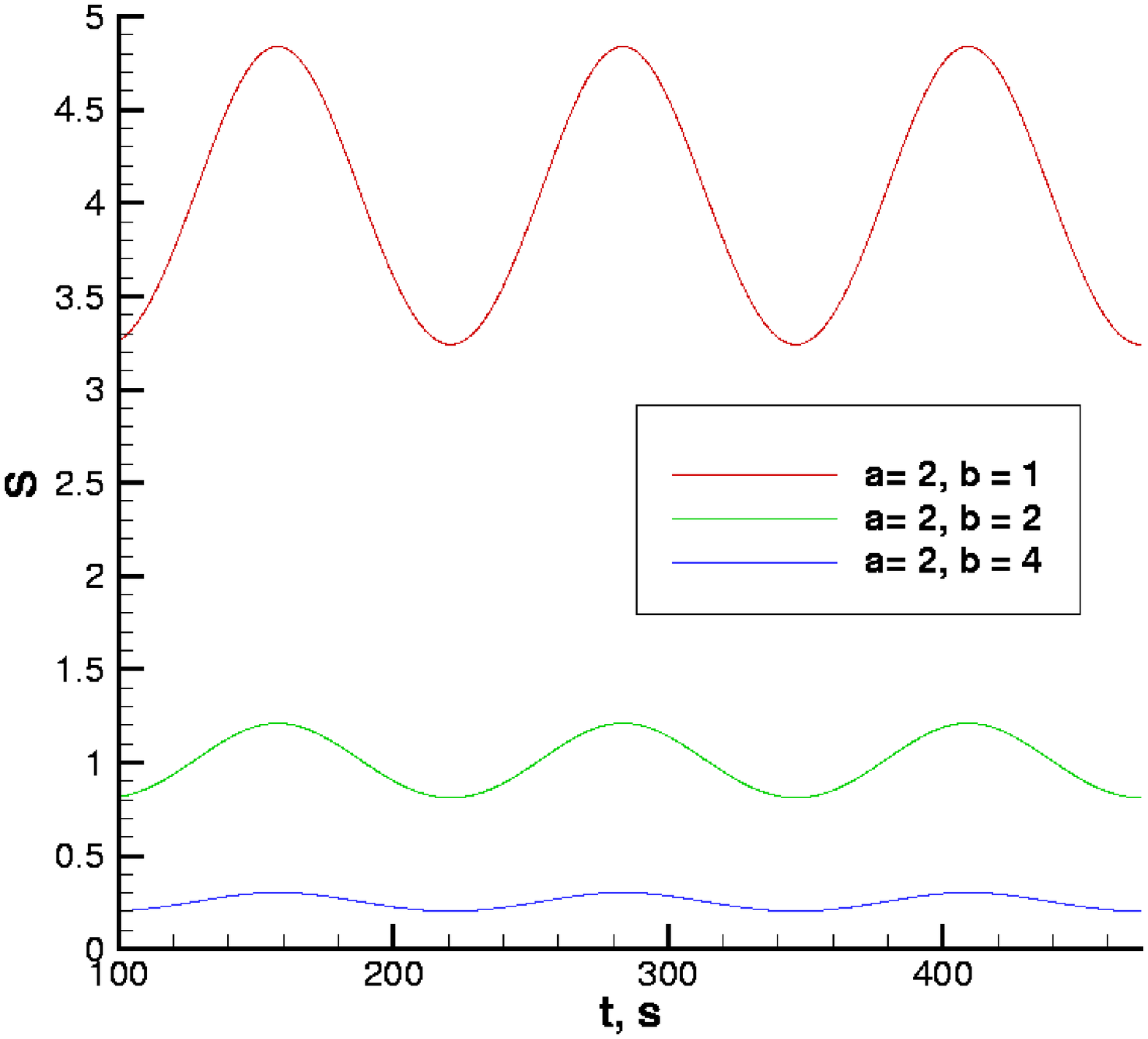}
\includegraphics[height=7.cm,clip=true]{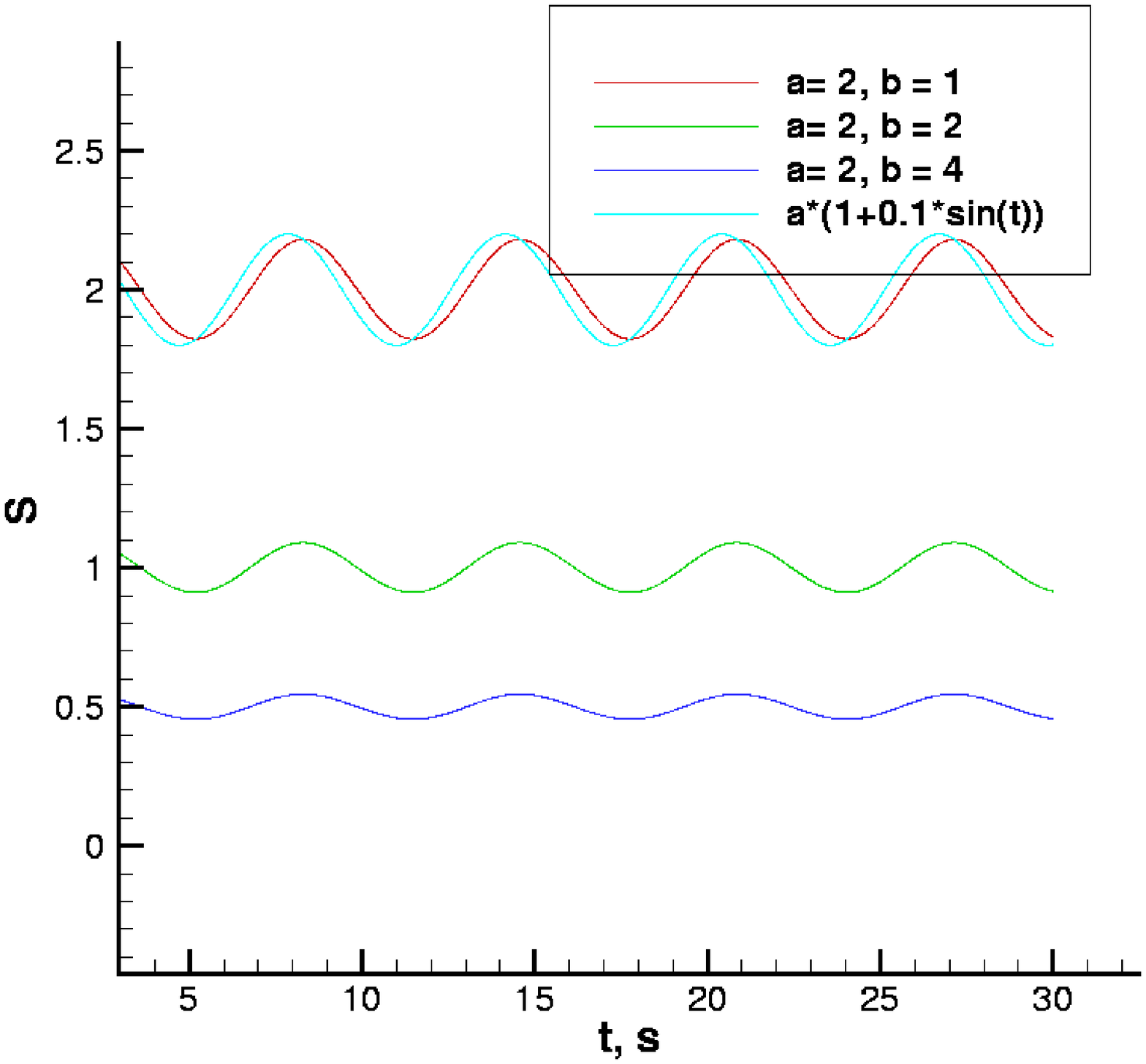}
\includegraphics[height=7.cm,clip=true]{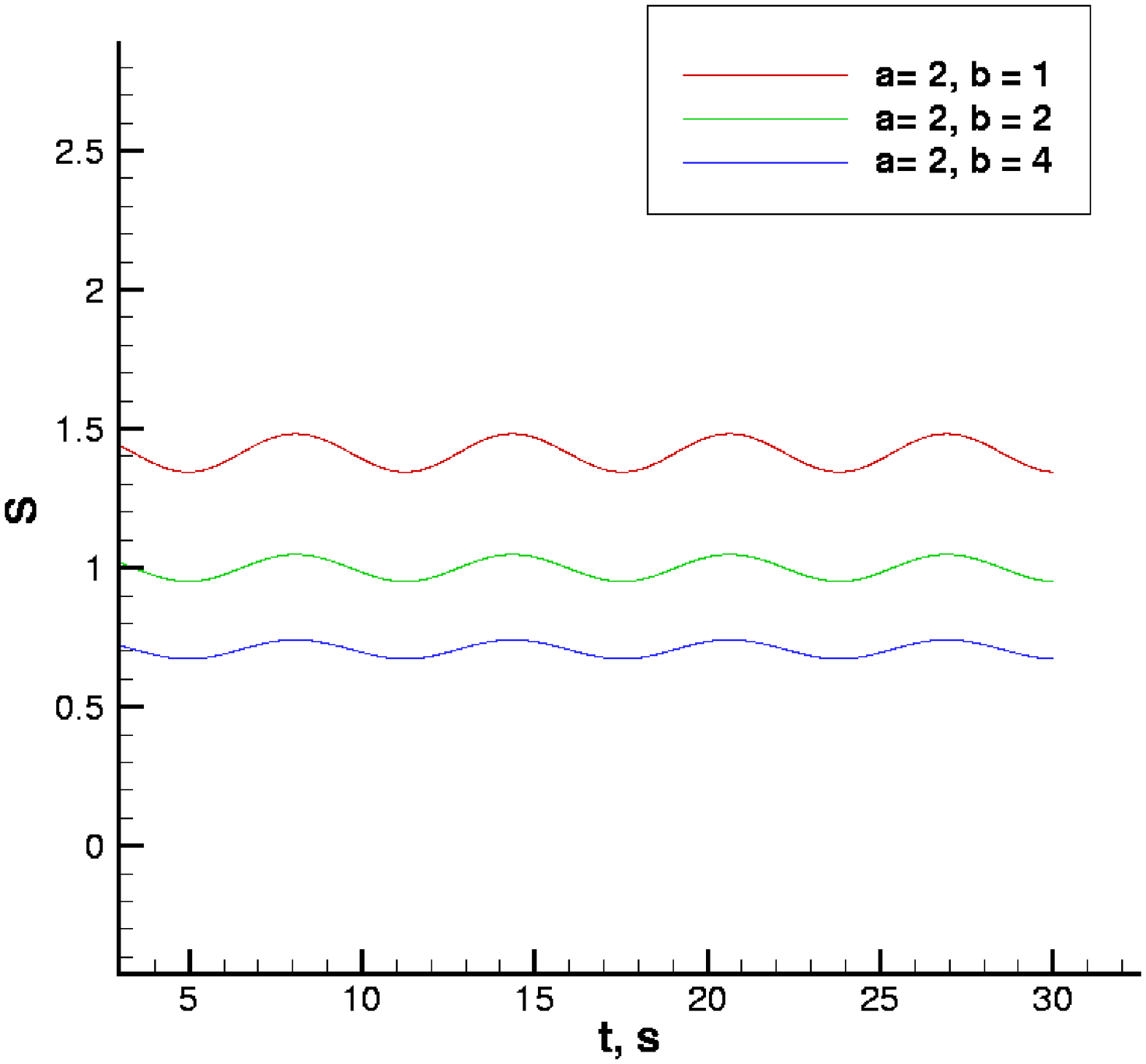}
\end{center}
\caption{Solution of Eq.~(\ref{e:dSdt_Sn}) with different $b$ and
  $n$'s. Top: $n = 1.5$; middle: $n = 2$; Bottom: $n = 3$.
  Creation = a(1. +  0.1sin($\omega$t))S for all cases. }
\label{f:St_diff_nb}
\end{figure}

Let us now consider a time-dependent perturbation, $\Delta a(t)$, and ask if the above properties
still obtain under this generalization. Assume
\begin{equation} \label{e:dela}
  \Delta a(t) = \Delta a_0\sin(\omega t)\,,
\end{equation}
where $\Delta a_0$ and $\omega$ are the amplitude and frequency of the
perturbation, respectively. We have obtained numerical solutions for $a =
2$ and $\Delta a_0 = 0.2$. Three different values of $n$ and $b$ (equivalently, the laminar flame speed) were
examined to determine the dependence of the equilibrium solution, $S_e$, and
the deviation from it, $\Delta S$, on $n$ and $b$.  
Fig.~\ref{f:St_diff_nb} shows these solutions. The dependences
of $S_e$ and $\Delta S$ on $n$ and $b$ described by Eqs.~(\ref{e:Se})
and~(\ref{e:delSelim}) remain the same under the generalization of perturbation. 
Specifically, for a given $a$ and $\Delta a_0$, we always obtain 
\begin{equation} \label{e:prpty1}
  S_e = \left(\frac{a}{b}\right)^\frac{1}{n-1}
\end{equation}

\begin{equation} \label{e:prpty2}
  \Delta S = \frac{\Delta a_0}{(n-1)a}\left(\frac{a}{b}\right)^\frac{1}{n-1}. 
\end{equation}
With these two functional dependences firmly established, we are left with determining the 
value of $n$ in (\ref{e:dSdt_Sn}) to completely specify the
governing equation of the flame surface area evolution.

As was demonstrated 
in Sec.~\ref{s:S_evol}, both the time-averaged value and the standard
deviation of the flame surface area is proportional to $1/D_l$.
Eqs.~(\ref{e:prpty1} and~(\ref{e:prpty2} then imply
that $n$,  in Eq.~(\ref{e:dSdt_Sn}) is indeed equal to 2. (A new subgrid model can then be developed based on Eq.~(\ref{e:dSdt}) of the flame surface area evolution. Details of the new model will be discussed in a forthcoming paper.) 
Therefore, we restrict our immediate discussion to the case $n = 2$. Note that we assume the creation
coefficient, $a$, and its deviation, $\Delta a_0$, are
constant for different laminar flame speed, $D_l$'s. We will provide evidence that this is a reasonable assumption in 
the following section.

In case $n = 2$,  Eqs.~(\ref{e:prpty1}) and~(\ref{e:prpty2}) yield
\begin{equation} \label{e:prpty1_n2}
  S_e = \frac{a}{b}\,,
\end{equation}

\begin{equation} \label{e:prpty2_n2}
  \Delta S = \frac{\Delta a_0}{b}. 
\end{equation}

The amplitude of the variation of the flame surface area, $S$, in the
steady state is proportional to the amplitude of the variation of the
perturbation, $\Delta a_0$.  Intuitively, there should exist a
relaxation time scale, $\tau_r$, associated with the response of $S$
to a perturbation in the creation coefficient, $a$.
For $a =$ const, the solution of Eq.~(\ref{e:dSdt_Sn}) is

\begin{equation} \label{e:Sn2exp}
    S = \frac{S_{0}S_{e}}{S_ee^{-at} + S_0(1-e^{-at})}\,,
\end{equation}

and can be also written as

\begin{equation} \label{e:Sn2}
    \frac{a-bS}{bS} = \frac{S_e-S}{S} = \frac{S_e-S_0}{S_0}e^{-at}\,,
\end{equation}

where $S_0 = S(t = 0)$. Therefore, the relaxation time scale $\tau_r$
is determined solely by the value of the coefficient $a$ itself.  For
specificity, let us assume $\tau_r = 1/a$.  Note that this relaxation
time scale does not depend on $b$ (or, equivalently, the laminar flame
speed).  We can see from Fig.~\ref{f:St_diff_nb} that, for the
perturbation defined in Eq.~(\ref{e:dela}), there is no phase
difference between cases with different laminar flame speed. The
difference in laminar flame speed affects only the steady state value
of $S$ and the deviation from it.

\begin{figure}[ht]
\begin{center}
\includegraphics[height=7.cm,clip=true]{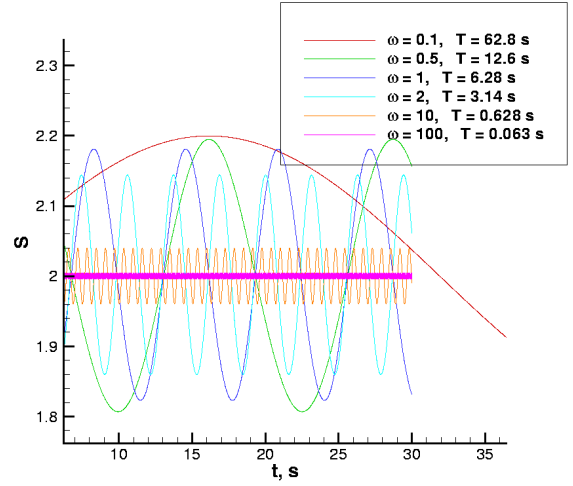}
\end{center}
\caption{Solution of Eq.~(\ref{e:dSdt_Sn}) for several values of $\omega$. $a$
  =2, $b$ =1, and $\Delta a_0$ = 0.2 for all cases. Note the
  percentage variation in flame surface area, $S$, is close to
  10$\%$ only when the period of excitation, $T$, is
  small compared to the relaxation time scale $\tau_r = 1/a$ = 0.5 s. }
\label{f:St_diffomg}
\end{figure}

Given that it takes some time for the flame surface to respond to a
change in the flow conditions, we might expect the
variation of the flame surface area to be different when the
frequency of the excitation, $\omega$, in Eq.~(\ref{e:dela}) changes.
To examine the response of the flame surface area to excitations of
different frequencies, a numerical experiment was performed with $a = 2$, $b =
1$ and $\Delta a_0 = 0.1a = 0.2$. The frequency of the excitation is
varied from 0.1 to 100 s$^{-1}$. Combining Eqs.~(\ref{e:prpty1_n2})
and~(\ref{e:prpty2_n2}), we obtain
\[
\frac{\Delta S}{S_e} = \frac{\Delta a_0}{a},
\]
{\it i.e.}, the percent variation in the  flame surface area  
is equal to the percent variation of the creation coefficient. It is clear from 
Fig.~\ref{f:St_diffomg} that only when the frequency of excitation
is low ($\omega < 1$) is the amplitude of the variation in $S$ 
as large as  the percentage variation of the creation
coefficient, {\it i.e.} $\sim$10$\%$. When the frequency of excitation is large ($\omega > 1$),
the flame surface becomes less responsive and the percentage by which
the flame surface area varies decreases to below 10$\%$. The relaxation time scale in this
case is $\tau_r = 1/a$ = 0.5 s. The period of excitation, $T$, is 6.28 s for an excitation 
frequency of 1 s$^{-1}$. Therefore, the period of excitation needs to
be about ten times larger than the relaxation time scale for the flame
surface to ``fully" respond to the excitation.

\subsection{Flame Surface Creation and Destruction}\label{s:cd}
After determining the form of the governing equation of the flame surface area 
evolution, let us now attempt to determine the parameters in the equation for the
actual physical system. 

\begin{figure}[ht]
\begin{center}
\includegraphics[height=7.cm,clip=true]{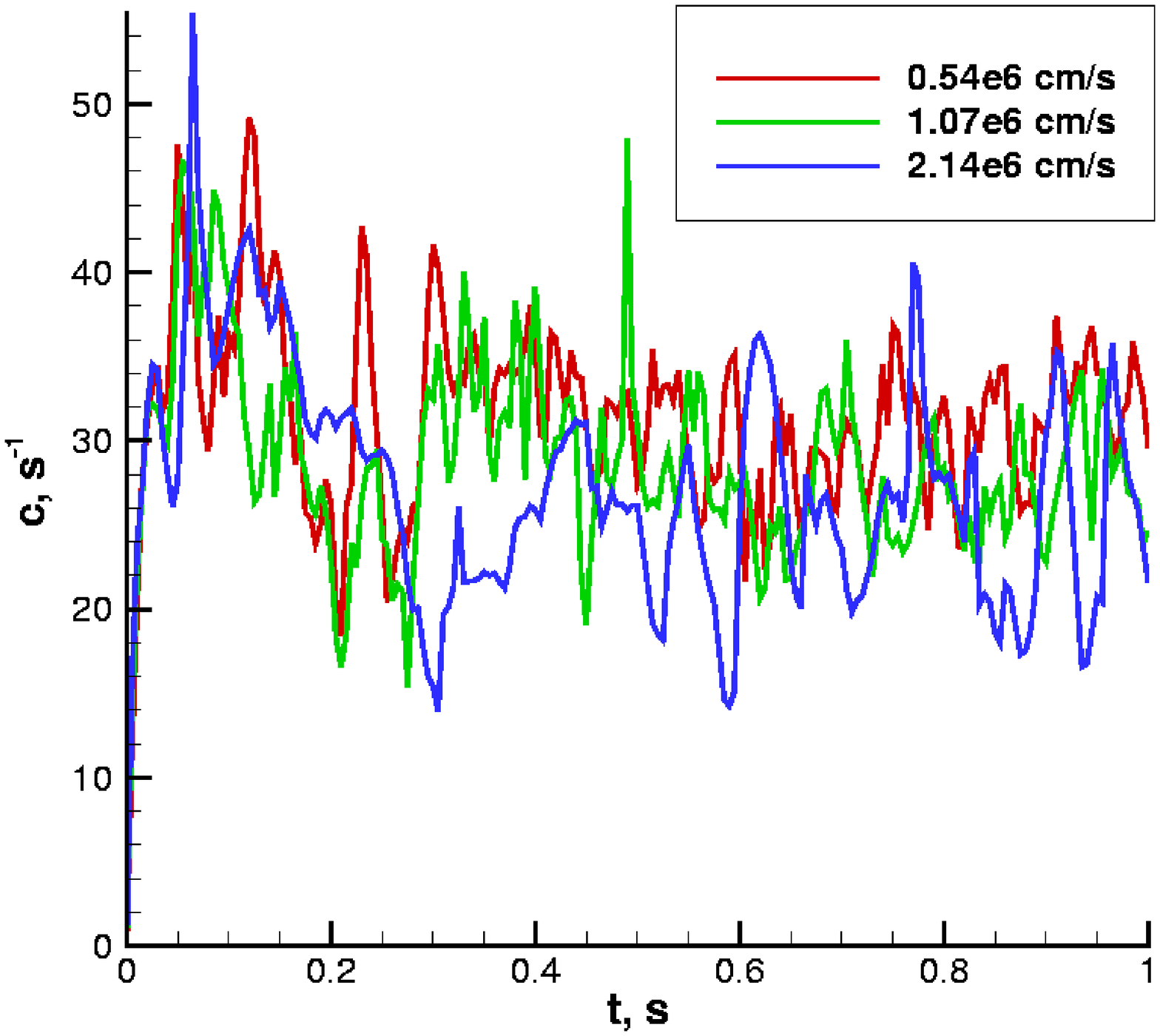}
\includegraphics[height=7.cm,clip=true]{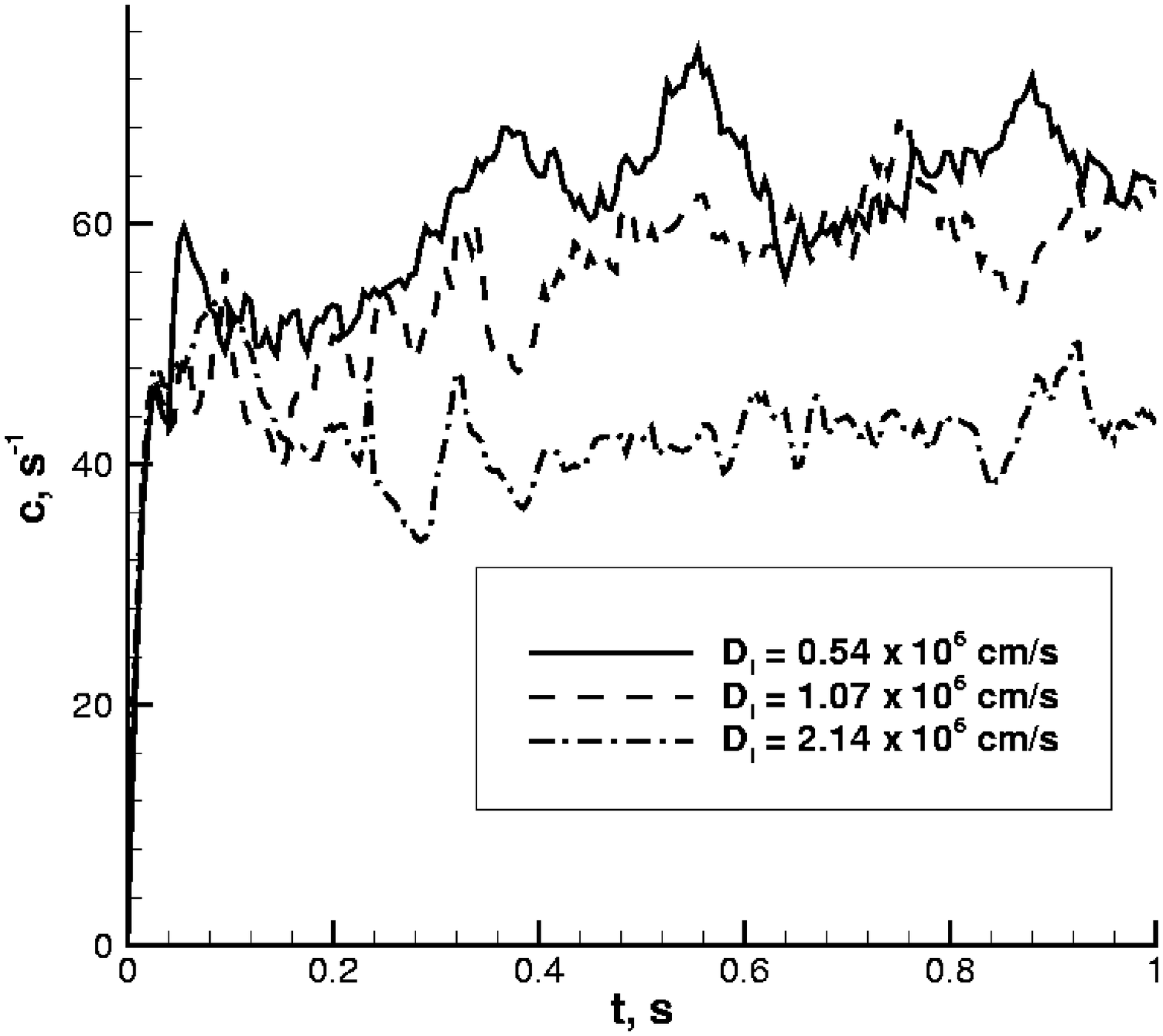}
\end{center}
\caption{The temporal evolution of the surface creation in 2D (left) and
  3D (right) in models DH, R1 and D2 and their counterparts in 2D. Note the
  similar time averaged value of $c$ at different laminar flame speeds.}
\label{f:c_Dl}
\end{figure}

As introduced at the beginning of Sec.~\ref{s:dynamics}, the evolution
of the creation coefficient, $c$, is followed in terms of its average 
value, $\int_\Omega cdV/\int_\Omega dV$, where $\Omega$ is
the total volume containing flame surface, {\it i. e.} all of the computational 
cells that contain flame surface.
Fig.~\ref{f:c_Dl}  

\begin{figure}[ht]
\begin{center}
\includegraphics[height=7.cm,clip=true]{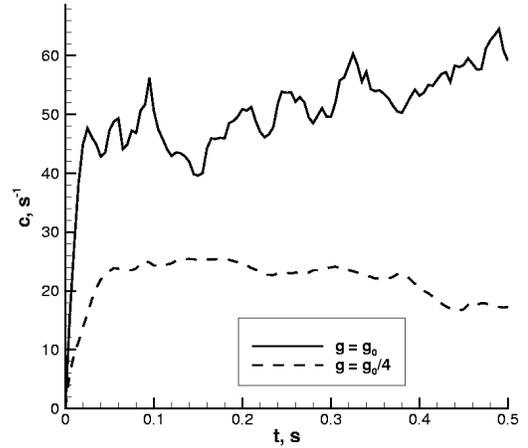}
\end{center}
\caption{The temporal evolution of the surface creation coefficient in
  3D models with nominal (solid) and 4 times lower gravity (dashed).}
\label{f:c_g}
\end{figure}

shows the evolution of the average creation coefficient for models DH,
R1 and D2. One can see that $c$ varies around $30$ s$^{-1}$ in 2D and
$60$ s$^{-1}$ in 3D at different laminar flame speeds.\footnote{Note
the lower creation coefficient in the D2 run, where the flow field is
not quite as turbulent as the other runs ({\it cf}
Figs.~\ref{f:flm_2d_Dl} and~\ref{f:flm_Dl_3d}).}  As discussed in the
previous section, the reciprocal of the creation coefficient, $a = c$,
is the relaxation time scale ($\tau_r$) of flame response to a change
in the flow conditions.
The relaxation time scale is $\approx 0.033$ and $0.017$\, s, in
these models for 2D and 3D, respectively, while the RT time scale,
$\tau_{RT} = \sqrt{L/Ag}$, is 0.05\, s.  In addition, $c \approx 50$
s$^{-1}$ in a model with nominal gravity while about twice smaller than
that in the model with gravity 4 times lower (see Fig.~\ref{f:c_g}).

Therefore, the reciprocal of the creation coefficient seems to also be
a measure of the RT time scale.  This is actually the physical
explanation for the independence of the relaxation time scale from the
laminar flame speed discussed in the previous section.

It is also interesting to note that the creation coefficient in 3D is about twice as 
large as that in 2D. It can be shown analytically that the creation coefficient for a uniformly
expanding sphere is twice as large as that of an expanding
cylinder. It appears this effect of dimensionality can be generalized to more generic shapes. 
The additional potential for surface area creation afforded by an additional dimension also 
serves to explain why our determined 
turbulent flame speeds are roughly  twice larger in 3D than in 2D ({\it cf} Figs.~\ref{f:Dtself}).

\begin{figure}[ht]
\begin{center}
\includegraphics[height=7.cm,clip=true]{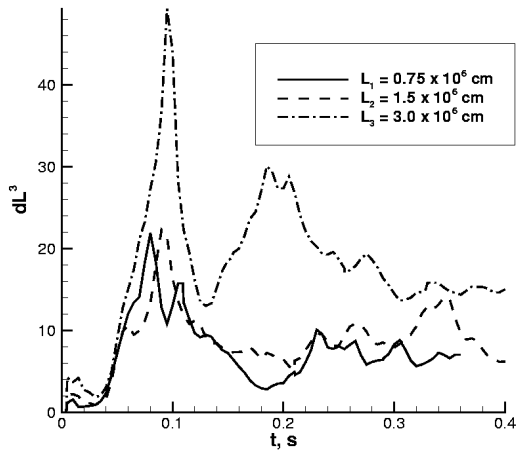}
\end{center}
\caption{Comparison of the time evolution of $dL^3$ 
  in models LH, R1 and L2, where $d$ is the destruction coefficient and $L$ is the lateral extent of
  the computational domain. Note that the destruction coefficient
  scales roughly as $1/L^3$. }
\label{f:d_V-1}
\end{figure}

\begin{figure}[ht]
\begin{center}
\includegraphics[height=7.cm,clip=true]{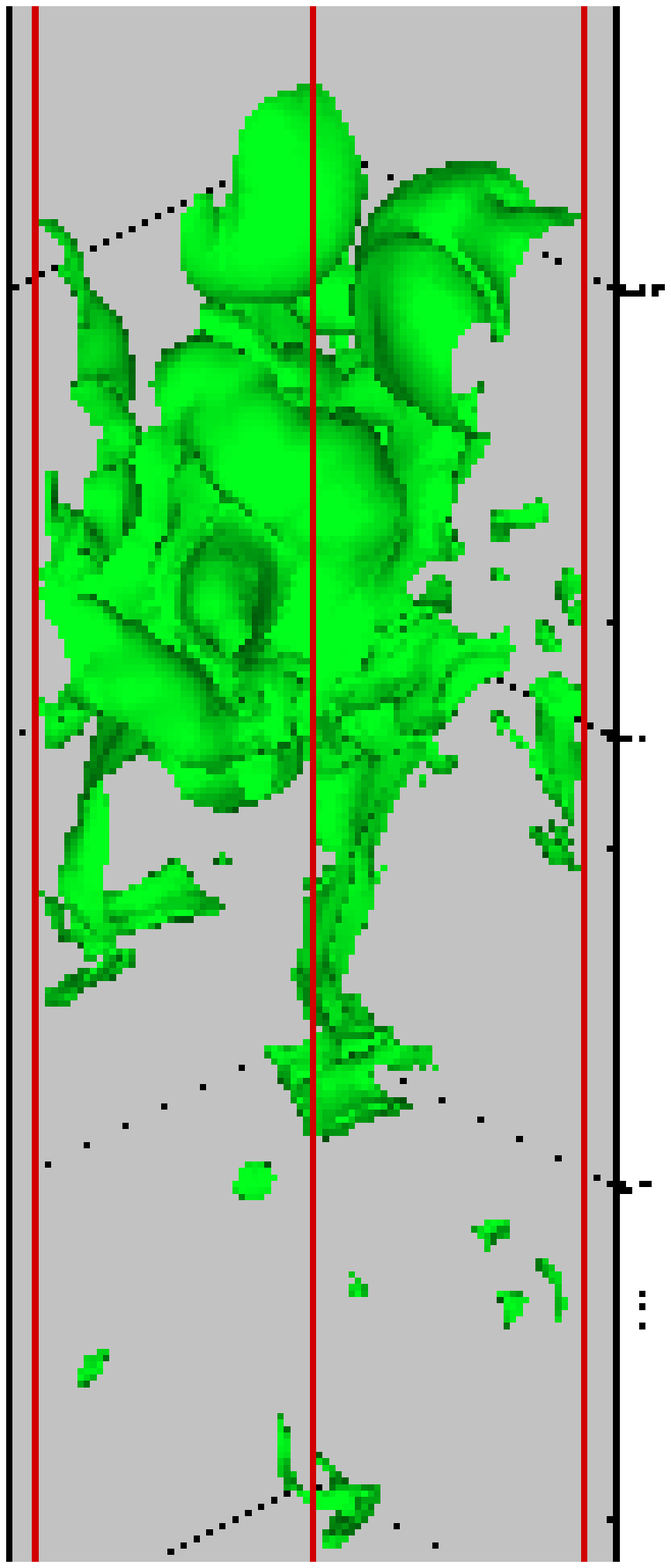}
\includegraphics[height=7.cm,clip=true]{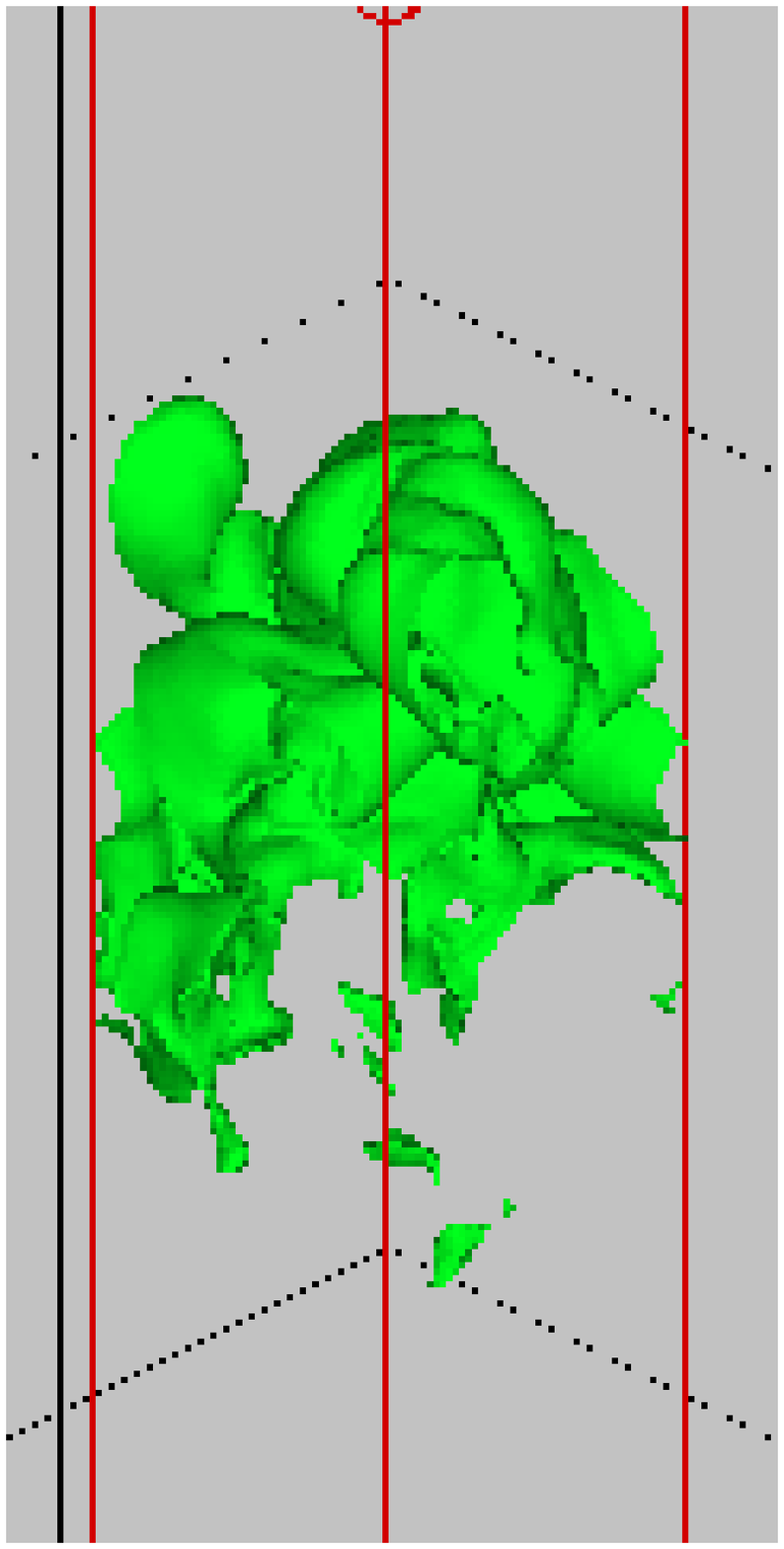}
\end{center}
\caption{Morphology of the turbulent flame surface in models with
  different domain sizes at late times.  (left) model R1: $L = 1.5
  \times 10^{6}$ cm; (right) model LH: $L = 0.75 \times 10^{6}$
  cm. Note that at late times the bubble size, $l$, becomes comparable
  to the size of computational domain, $L$.}
\label{f:V_bub}
\end{figure}

The numerical value of the surface destruction coefficient, $d$, can be
estimated from the simulation data using Eq.~(\ref{e:dSdt}). We
obtain the destruction term by subtracting the rate of change of the
flame surface area from the flame surface creation term.  $d$ is then obtained by dividing the
destruction term by $D_lS^2$.  Dimensional analysis indicates
that the coefficient $d$ must have the dimensions of specific volume. It 
is therefore reasonable to expect that $d$ scales as $1/L^3$, where $L$ is the lateral
extent of the computational domain. To
verify this result, we calculate three models of turbulent flame
evolution in boxes of different lateral dimension, each model having
lateral extent twice that of its predecessor (models LH, R1 and
L2). The evolution of the destruction coefficient in those runs is
shown in Fig.~\ref{f:d_V-1},

which clearly indicates that the
destruction coefficient scales with the inverse of the simulation
volume, {\it i.e.} $d \propto 1/L^3$. Given that the Rayleigh-Taylor
bubble volume is comparable to $L^3$ at late times
(Fig.~\ref{f:V_bub}), one can state that the 
destruction coefficient scales with the inverse of the characteristic Rayleigh-Taylor
bubble volume.

The following interpretation of destruction coefficient scaling with
the inverse of the Rayleigh-Taylor bubble volume can be offered:

Consider a box of size $L$ per dimension filled with segments of flame
surface. Let us assume that the segments are parallel to each other
and are separated by distance $\Delta$. Moreover, let us assume that
the neighboring flame segments are moving towards each other with a
constant speed, $D_l$. Then the surfaces are destroyed in collisions
at the rate

\[
\frac{S}{\Delta/D_l} = \frac{S^2}{S\Delta/D_l}\,.
\]
In this case, the destruction coefficient is

\[
 d = \frac{1}{S\Delta}\,,
\]
which can been seen as a measure of how much flame surface can be
packed into a given volume. The term $S\Delta$ can be identified with 
the bubble volume.



The fact that the destruction coefficient is inversely
proportional to the bubble volume, combined with dimensional analysis
of the surface area change rate equation strengthens our conclusion that the
surface destruction term is indeed proportional to $S^2$,
in agreement with the original proposal of \citet{khokhlov95}.
\subsection{Froude Number and the Self-Regulation Mechanism}\label{s:Fr}
The flame surface creation and destruction processes reflect the
interplay between the Rayleigh-Taylor instability and the tendency
of the burning process to smooth the flame's surface. This
interplay  has been clearly
illustrated in Fig.~\ref{f:flm_Dl_3d}
where we show the flame surface in three models obtained with
different laminar flame speeds. Model DH (left panel in
Fig.~\ref{f:flm_Dl_3d}) has the lowest laminar flame speed and
develops a flame surface which is substantially more convoluted than
in the remaining two cases where the laminar flame speed is higher
(models R1 and D2, middle and right panel in Fig.~\ref{f:flm_Dl_3d},
respectively). The flame surface is relatively smooth in high laminar speed 
model D2.  This
behavior is well characterized by the Froude number,
\[
Fr = \frac{D_l^2}{g L}\,,
\]
which is the only parameter required to completely characterize the
behavior of our model flames provided the density is
constant. In models with higher Froude number the flame surface
smoothing due to burning is stronger, resulting in a less convoluted
flame surface. Conversely, in models with lower Froude number, the
flame surface is strongly convoluted. The burning rate, however,
remains controlled by the self-regulation mechanism and is independent
of Froude number ({\it i.e.} independent of the laminar flame speed).

>From the point of view of flame surface creation
processes, systems characterized by smaller Froude number are expected
to have similar surface creation coefficients, independent of $D_l$ as
long as the RT time scale are similar (as seen in
Fig.~\ref{f:c_Dl}). However, the creation coefficient will become sensitive
to the laminar flame speed once the Froude number is such
that the flame 
speedup factor, {\it i.e.} the ratio of turbulent flame speed to the
laminar flame speed, becomes less than $\sim$10 for 3D. 

%
%
%
\subsection{Verification of the KSGS model}\label{s:verification}
\subsubsection{Numerical Convergence}
\begin{figure}[ht]
\begin{center}
\includegraphics[height=7.cm,clip=true]{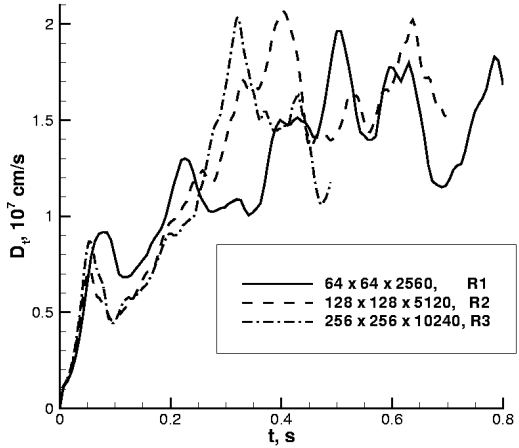}
\end{center}
\caption{Convergence study for the turbulent flame models. Shown is the temporal
  evolution of the turbulent flame speed, $D_t$, in models R1, R2 and R3.}
\label{f:dMbdt_conv}
\end{figure}

To examine the convergence behavior of our numerical flame models, we perform simulations at
three different spatial resolutions (models R1, R2 and R3).  Fig.~\ref{f:dMbdt_conv}

shows the evolution of the turbulent flame speed, $D_t$, in these
three models. At sufficiently late times the flame speed exhibits
variations of $\approx 15\%$ around a mean value of $\approx 1.6\times
10^7$\,cm/s, and the detailed behavior is remarkably similar in all three
cases. (Note that, due to limited computing resources, the highest
resolution calculation was stopped at $t\approx 0.5$\,s. Nevertheless,
even in this case the flame speed increase saturates at a similar
level as in the two lower resolution cases.) We also find that at any
given instant of time the total burned mass is similar in all three
models.

\begin{figure}[ht]
\begin{center}
\includegraphics[height=5.5cm,clip=true]{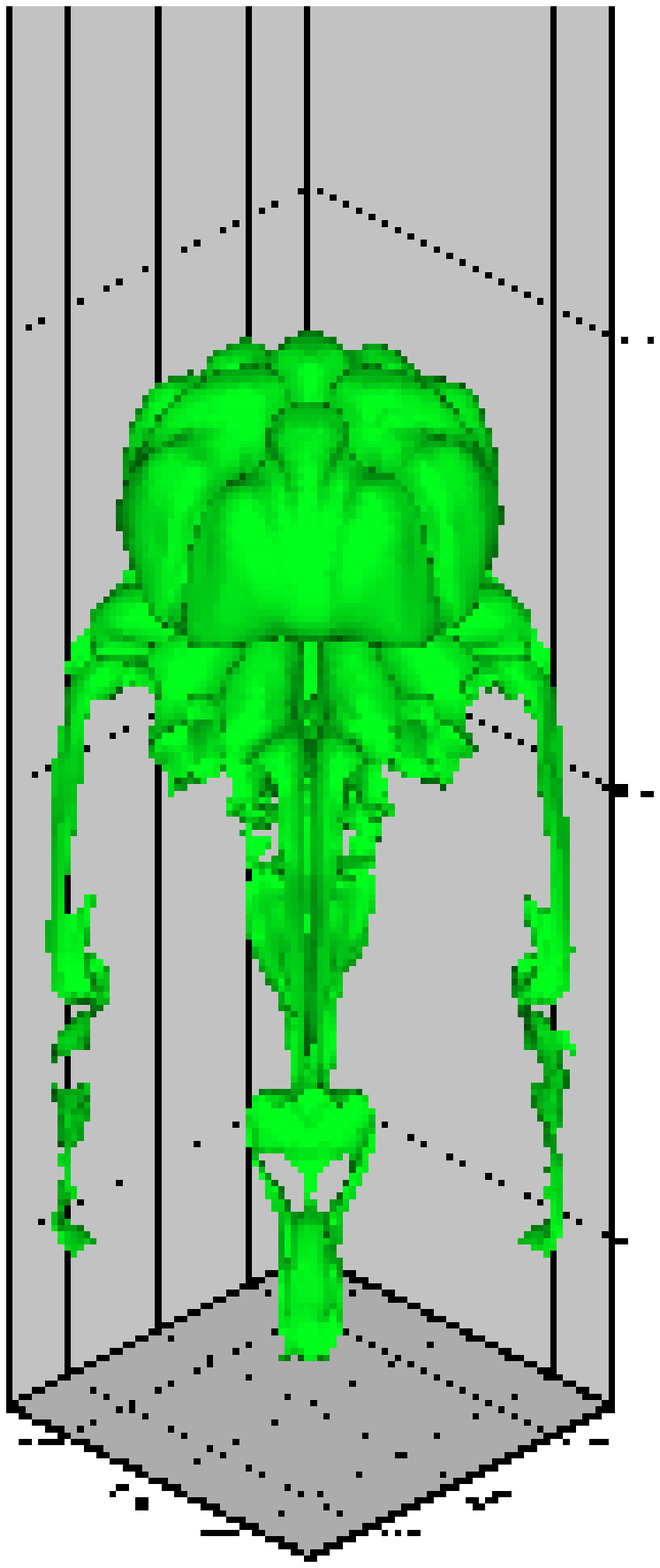}
\includegraphics[height=5.5cm,clip=true]{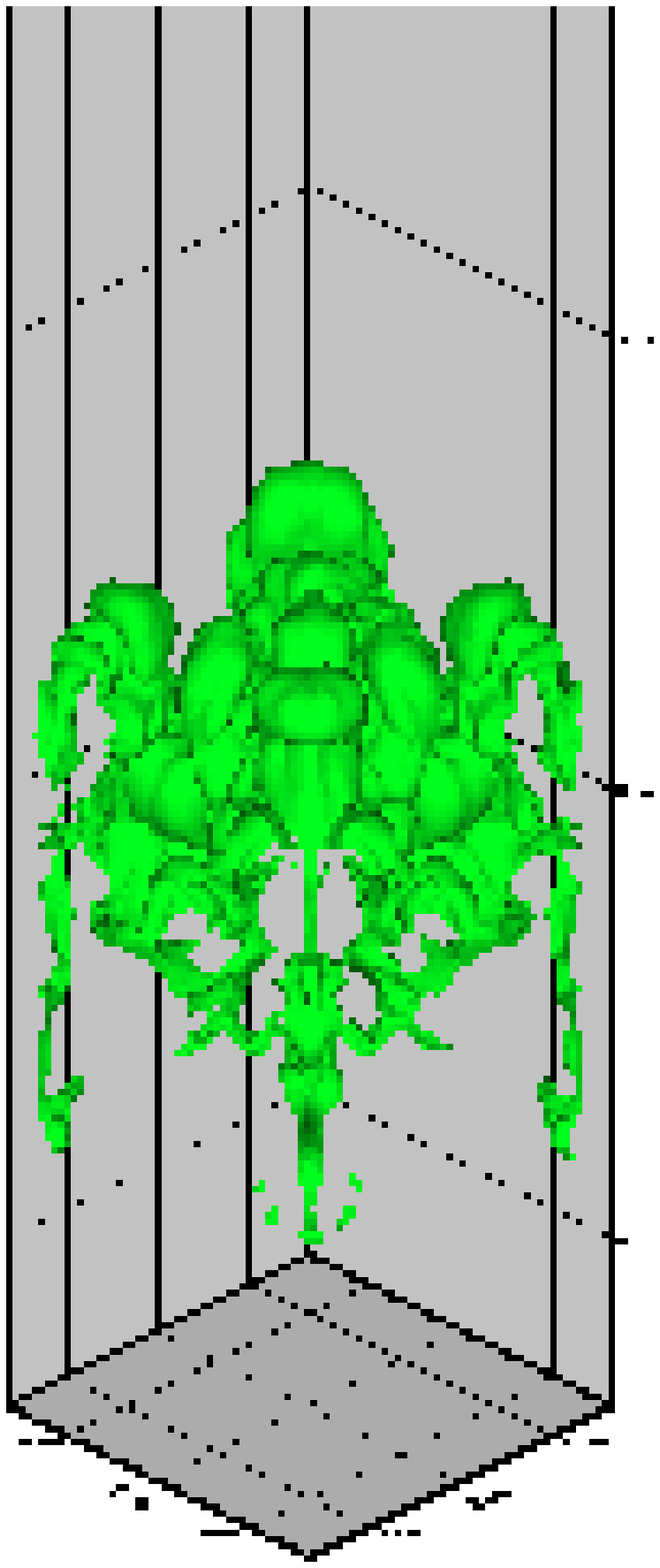}
\includegraphics[height=5.5cm,clip=true]{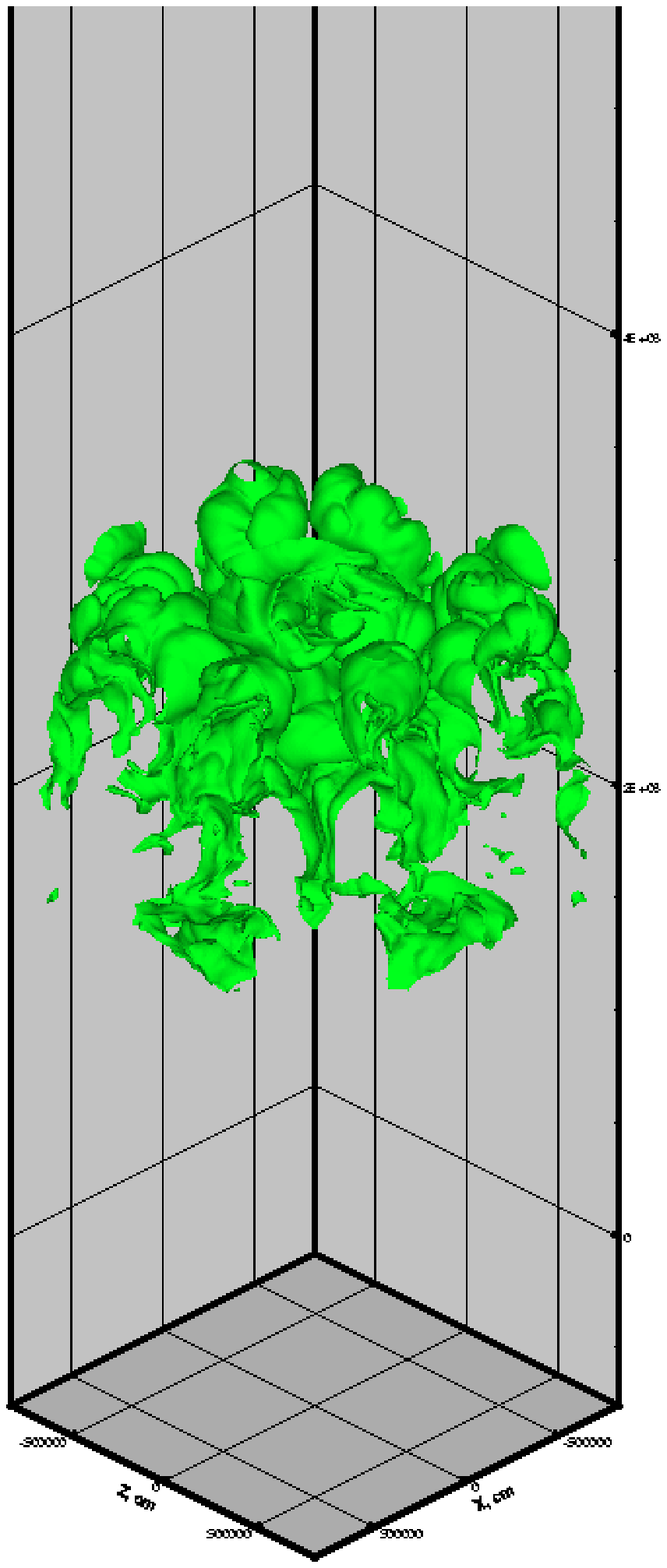}
\end{center}
\caption{Dependence of the morphology of the turbulent flame surface
  on numerical resolution. The surface of the flame is shown at $t\sim
  0.2$\,s in model R1 (left panel, $\Delta x=2.34\times 10^4$\,cm), R2
  (middle panel, $\Delta x=1.17\times 10^4$\,cm) and R3 (right panel,
  $\Delta x=5.86\times 10^3$\,cm).}
\label{f:S_l345}
\end{figure}

Due to the inviscid nature of the Euler equations, the flow structure
will depend on numerical resolution to a degree. This is illustrated
in Fig.~\ref{f:S_l345}

where we show the morphology of the flame surface in models R1, R2, and R3
obtained at increasing resolution (but having fixed Froude
number). One can notice a dramatic increase in the amount of
small-scale structure as the resolution is increased. However,
once the large scale characteristics of the flame
are captured by the numerical model, the small scale structure
emerging with increased numerical resolution does not change the global
properties of the evolution, as seen in the convergence of the
turbulent flame speed shown in Fig.~\ref{f:dMbdt_conv}.

\begin{figure}[ht]
\begin{center}
\includegraphics[height=7.cm,clip=true]{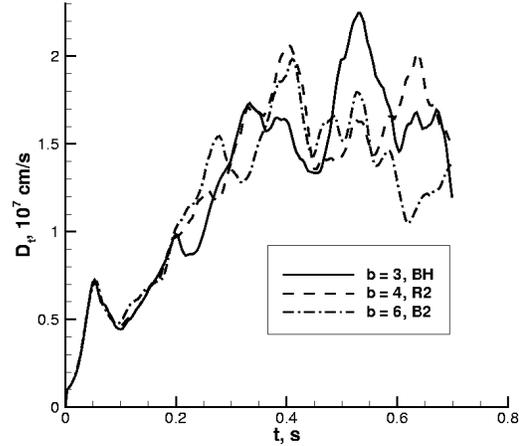}
\end{center}
\caption{Dependence of the turbulent flame speed on the nominal
numerical thickness of the flame front, $b$, in models BH, R2 and
B2. Note that in all cases the turbulent flame speed oscillates around
a similar average value.}
\label{f:Mb_b346}
\end{figure}

Our results might also be expected to depend on certain technical aspects of the
simulations.  For example, due to the finite thickness of the numerical flame
front  it is desirable to study the dependence of our results on
this numerical parameter.  For this purpose we obtained models BH and
B2, in which we adjust the nominal numerical flame thickness to,
respectively, 0.75 and 1.5 times the standard value used in model
R1. This is accomplished by changing the parameter $b$ from 4 to 3 (model BH)
or 6 (model B2). Figure~\ref{f:Mb_b346}

shows that at late times the turbulent flame speed in all three models
oscillates around a similar mean value with an amplitude entirely
consistent with that observed in the resolution study discussed above.

\begin{figure}[ht]
\begin{center}
\includegraphics[height=7.cm,clip=true]{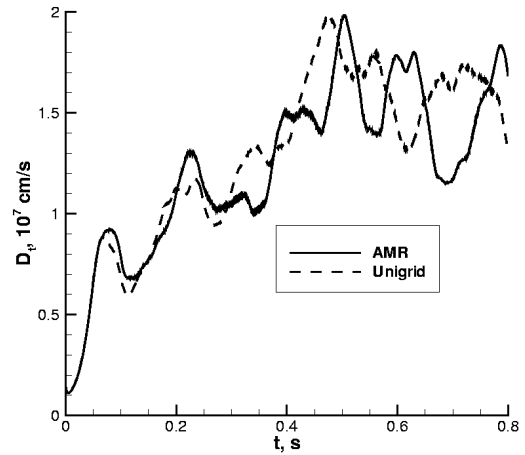}
\end{center}
\caption{Verification of the adaptive mesh refinement model of the
  turbulent flame. Temporal evolution of the turbulent flame speed in
  the AMR model is plotted along that obtained in the simulation with
  uniform grid. The two models produce statistically similar results.}
\label{f:uni_amr}
\end{figure}

Since our fiducial simulations make use of adaptive discretization, it is
necessary to verify that local nonuniformity of the computational grid
does not markedly influence the evolution of the
system. Figure~\ref{f:uni_amr}

shows the evolution of the turbulent flame speed in model R1 and its
counterpart calculation on a uniform mesh. The two models perfectly
agree at very early times ($t < 0.1$\,s) when the flame evolution
remains confined to the region initially refine. The two models begin to
differ during the initial transient phase when the flame structure
becomes progressively turbulent and the mesh in model R1 begins to
follow the evolving flame structure. However, in steady state, the
flame speeds in the two models saturate at similar levels. Also, we
find that the power spectrum of the time variation of the flame surface area
obtained in the model calculated on a uniform grid shows the same
salient features (location of peak power, concentration at long
timescales) as the AMR-enabled model.
\subsubsection{Examining the Self-Regulation Mechanism}\label{s:selfreg_01}
\begin{figure}[ht]
\begin{center}
\includegraphics[height=7.cm,clip=true]{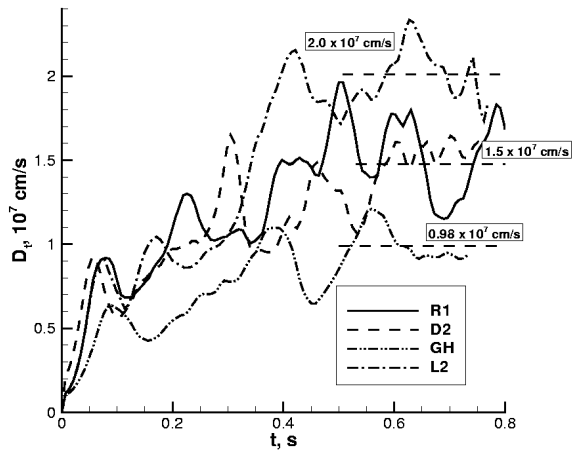}
\end{center}
\caption{Verification of the KSGS model of RT-driven turbulent
  flames. The time evolution of the turbulent flame speed is shown for models
  R1 (reference model), D2 (twice laminar flame speed), GH (half gravitational acceleration), and L2 (twice larger domain
  size). Time averaged turbulent flame speeds in the steady-state regime are within
  $10\%$ of the values predicted by Eq.~(\ref{e:AgL}).}
\label{f:AgL_veri}
\end{figure}

One of the main goals of this study is to verify the KSGS
subgrid-scale model proposed by \citet{khokhlov95}. The KSGS model
considers turbulence driven by the Rayleigh-Taylor instability on
scales comparable to the grid resolution. In this model, the
instantaneous turbulent flame speed, $D_t$, is
\begin{equation} \label{e:AgL}
  D_t = 0.5 \sqrt{A g L}\,, 
\end{equation}
where $A = (\rho_0-\rho_1)/(\rho_0+\rho_1)$ is the Atwood number,
$\rho_0$ and $\rho_1$ are the densities ahead and behind the flame
front, respectively, $g$ is the gravitational acceleration, and $L$ is
the turbulent driving scale, usually set to one or two computational
cell widths in large-scale simulations making use of the KSGS model \citep{gamezo+03}. A number of models were calculated to verify the applicability of
Eq.~(\ref{e:AgL}) for different $g$
and/or $L$ . The results are shown in Fig.~\ref{f:AgL_veri}.

The time-averaged steady-state turbulent flame speeds
are entirely consistent with the predictions of Eq.~(\ref{e:AgL}).

Another aspect of the verification of the KSGS model is to confirm the
independence of the self-regulation mechanism from details of the flame 
microphysics, {\it i.e.} on the laminar flame speed. To verify this,
models DH, R1, and D2 were obtained for the same initial conditions
and resolutions, but using laminar flame speeds differing by a factor
of 2.  The long-term evolutions of
the turbulent flame speeds in these 
models are shown in Fig.~\ref{f:Dtself}, which clearly illustrates the
expected independence of the turbulent and laminar flames speeds.
\section{Summary}\label{s:summary}
We present the results of an extensive study of the evolution of
thermonuclear flames in periodic domains under the influence of a
constant gravitational field. The flame surface is Rayleigh-Taylor
unstable, leading to a growth of initial perturbations.  We follow the
flame's development through an extended initial transient phase and
find a well-defined steady-state regime. We verify that the evolution
of  RT-unstable flames is self-regulated, {\it i.e.} the turbulent
flame surface evolution in steady state is independent of the details
of the incorporated microphysics.

The evolution of the flame surface was examined. The functional dependence of 
flame surface destruction suggested by (Khokhlov 1995) is proven and
the properties of the governing equation of the surface evolution are
discussed. 
The flame surface creation coefficient is found to be closely related
to the RT time scale and the flame
surface destruction strength is found to vary as $1/L^3$, where $L$ is
the turbulent driving scale.
The flame surface creation and destruction processes reflect the
interplay between the Rayleigh-Taylor instability and the flame's
tendency to smooth its surface. We find this relationship can be well
characterized by the Froude number.

We find that the turbulent flame speed obeys a $\sqrt{AgL}$ scaling
law for Froude numbers $\ll 1$. Also, we are able to identify
semi-periodic temporal variations in the flame evolution in steady
state and associate them with the turnover time of the largest eddies;
no significant amount of power on shorter time scales is found,
indicating the evolution is governed by large-scale flow characteristics. 

We discussed in some detail the mechanism of vorticity generation,
shedding, and transport in the vicinity of the flame surface. We
find that the interaction between induced vorticity and the flame surface
plays an important role in the flame evolution, leading to bubble
``wobbling'' on timescales comparable to the eddy turnover time. As
the bubbles rise, vorticity is advected downstream into the
bulk and forms an homogenized, mixed turbulent layer. It is in this
layer where most of the flame surface destruction takes place.
Although there exisst noticeable differences between flame dynamics in
2D and 3D, the vorticity production and advection, as well as the
interaction between vortices and the flame surface appear quite
generic. 

There are several possible future directions of research emerging from
our study. Our convergence study indicates that the self-regulation
mechanism is relatively insensitive to the numerical resolution,
provided the driving scales are adequately resolved. However, at the
highest tested resolutions, we exhaust our current computing resources
and are unable to investigate evolution with the Gibson scale fully
resolved. Such investigation, bridging large and small scales in one
model, is highly desirable. A full 3D implementation and tests
including large-scale Type Ia simulations with a new subgrid model will
also be conducted.
\acknowledgements This work is supported in part by the U.S.\
Department of Energy under Grant No.\ B523820 to the Center for
Astrophysical Thermonuclear Flashes at the University of Chicago.
This research used resources of the National Energy Research
Scientific Computing Center, which is supported by the Office of
Science of the U.S. Department of Energy under Contract No.\
DE-AC03-76SF00098, and of the Teraport project at the University of
Chicago through National Science Foundation Grant No.\ 0321253. We
would also like to thank Timur Linde for implementing the marching cubes module  in
FLASH and Janusz Kalu\.zny for making implementation of the CLEAN
algorithm available to us.

\begin{figure}[h]
\begin{center}
\includegraphics[height=6.cm,clip=true]{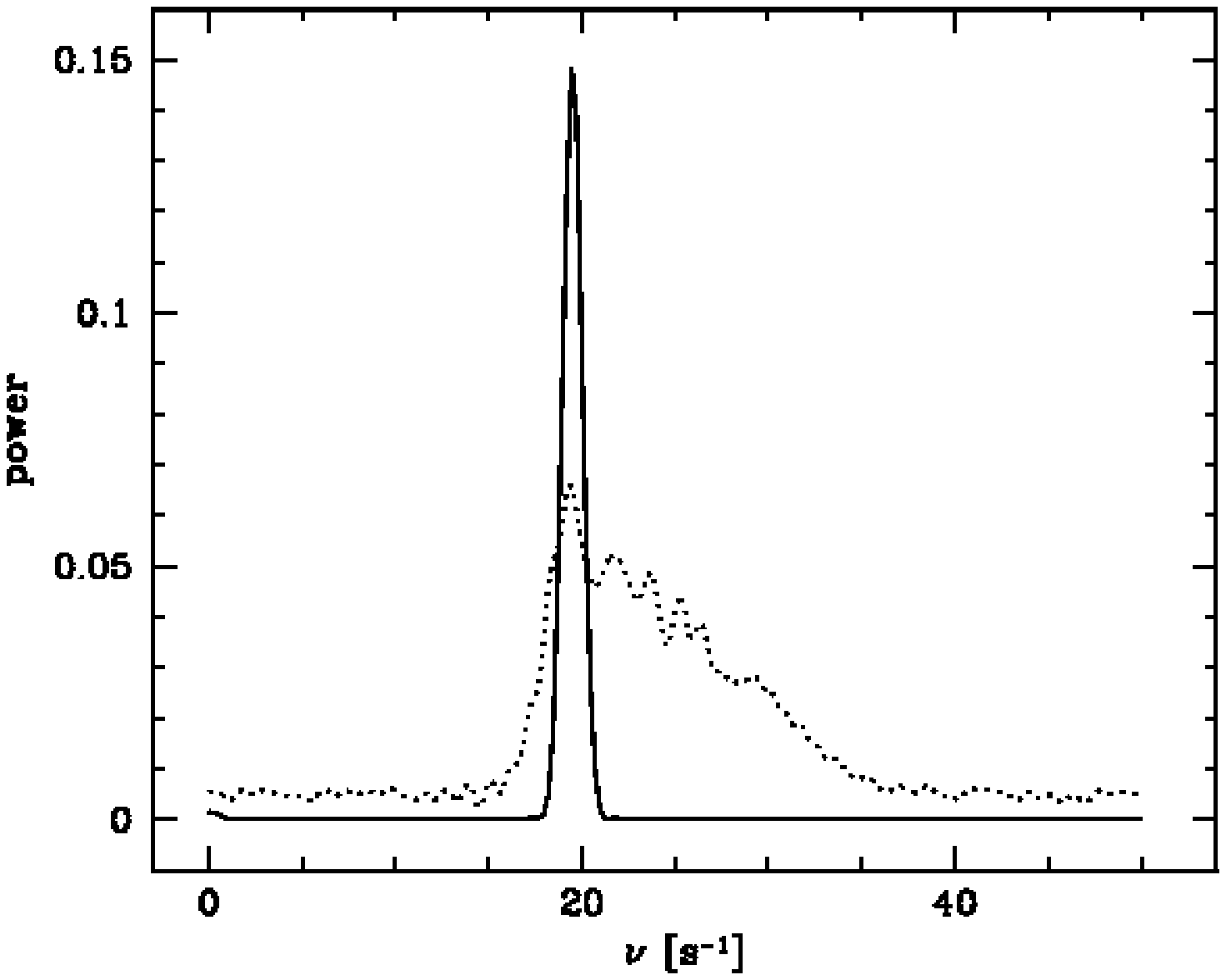}
\end{center}
\caption{CLEAN power spectrum of trial time series with single frequency without (solid) and with (dotted) frequecy drift. See text for details.}
\label{f:clean_veri}
\end{figure}

\appendix\label{s:appendix}
In order to verify the correctness and applicability of the CLEAN algorithm implementation,
we have generated and analyzed a set of trial time series.  The first test
time series consists of single frequency signal with $\nu = 19.5$~s$^{-1}$.
Figure~\ref{f:clean_veri}

shows that this single frequency is clearly identified by
CLEAN (solid line). We can extend this trivial example to consider the effect of frequency drift on
our ability to identify the spectrum.  In a second test the time variable is 
transformed using a simple power-law, $t^{(0.9+0.2\times(t/t_e))}$,
where $t_e$ is the final time. The resulting power spectrum is shown
in Fig.~\ref{f:clean_veri} (dotted line). The power now occupies a much wider range of frequencies, consistent with the assumed
drift of the base frequency.   We posit that such a drift around a
base frequency occurs in the complex interaction between the RT-unstable
flame with its own induced turbulent velocity field.  This leads to a substantial leakage of
power to frequencies neighboring the characteristic RT (turnover) frequency.  

\begin{figure}[h]
\begin{center}
\includegraphics[height=6.cm,clip=true]{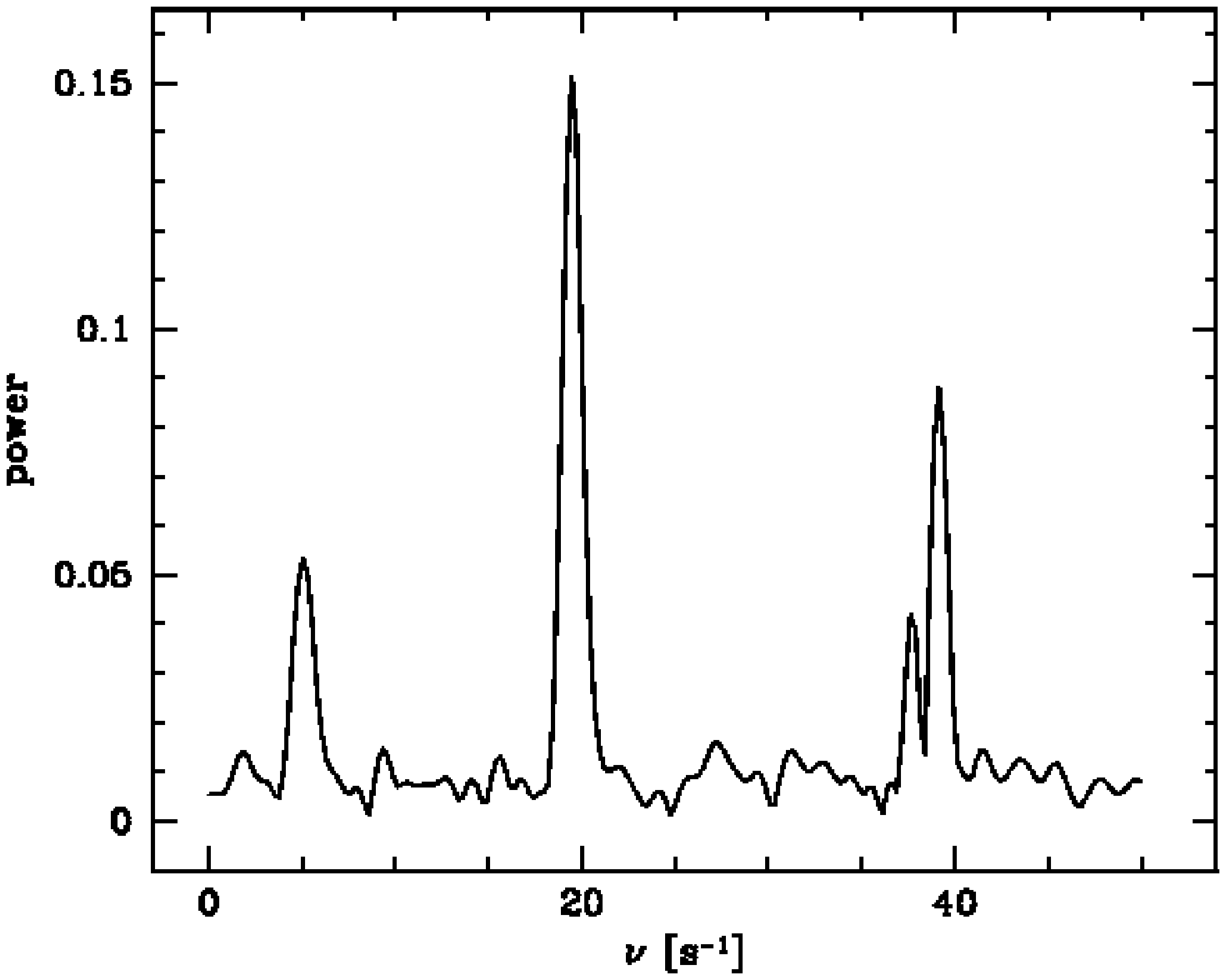}
\end{center}
\caption{CLEAN power spectrum for signal with multiple, closely-spaced,
  and aliased frequencies in the presence of significant white noise. Note
  that all major component frequencies are clearly identified with the 
  exception of the two low frequency signals (power peak around $\nu = 5$
  s$^{-1}$).}
\label{f:clean_veri_mt}
\end{figure}

In another verification test, the trial signal consists of multiple discrete 
frequencies superposed on a white noise signal:
\[
   s(t) = A_0 r_0(t)
   \begin{array}{l}N\\\sum\\i=1\end{array}A_i[1+r_i(t)]\cos(\pi\nu_i(2t+10^{-2}r'_i(t))),
\]
where $N = 4$, $r_{\{0,i\}}$ and $r'$ are time-dependent random numbers
between -1 to 1, $(A_i,\nu_i)$ (chosen to be are (0.1,4), (0.1,5), (0.3,19.5), (0.1,38)) are the amplitudes and matched frequencies of 
the discrete signals, respectively, and the amplitude of the white noise is $A_0=0.1$. The time series consisted of 100 
equally spaced points in time between 0 and 1. 
Despite the fact that (a) some frequencies are closely spaced and/or aliased,
(b) the amplitude of white noise is comparable to that of the composite
signals, and (c) small random perturbations have been added in the time
domain, most all the informational content in the signal is 
identified by the CLEAN method (Fig.~\ref{f:clean_veri_mt}).

Only in the case of the two lowest closely-spaced frequencies ($\nu=$4,5) does the method fail to clearly delineate the
separate signals.  Even in this case the power peak is relatively broad, suggesting the presence of distinct signals in
the region.

\end{document}